\documentclass[preprint,prd,aps,showpacs,showkeys,nofootinbib]{revtex4}
\usepackage{amssymb}
\usepackage{}
\usepackage{amsmath}
\allowdisplaybreaks[4]
\usepackage{multirow}
\usepackage{graphicx}
\usepackage{float}

\begin{document}

\preprint{}

\title{Rare top quark decays in the minimal R-symmetric supersymmetric standard model}

\author{Ke-Sheng Sun$^a$\footnote{sunkesheng@126.com, sunkesheng@bdu.edu.cn}, Zhi-Chuan Wang$^{b,c,d}$\footnote{404275079@qq.com}, Xiu-Yi Yang$^{e}$\footnote{yxyruxi@163.com},Tie-Jun Gao$^{f}$\footnote{tjgao@xidian.edu.cn}, Hai-Bin Zhang$^{b,c,d}$\footnote{hbzhang@hbu.edu.cn}}

\affiliation{$^a$Department of Physics, Baoding University, Baoding 071000, China\\
$^b$Department of Physics, Hebei University, Baoding 071002, China\\
$^c$Key Laboratory of High-Precision Computation and Application of Quantum Field Theory of Hebei Province, Baoding 071002, China\\
$^d$Research Center for Computational Physics of Hebei Province, Baoding 071002, China\\
$^e$School of science, University of Science and Technology Liaoning, Anshan 114051, China\\
$^f$School of Physics, Xidian University, Xi'an 710071, China}

\begin{abstract}
The one-loop contributions to the flavor-changing neutral current decays of the top quark into a light quark and a gauge boson or Higgs boson, namely $t\rightarrow qV,qh$, with $q$ being $u$ or $c$, and $V$ being $\gamma$, $g$, or $Z$, are analyzed in this study within the framework of the minimal R-symmetric supersymmetric standard model. The charginos in this model have been separated into two sets, i.e., $\chi$-chargino and $\rho$-chargino. The numerical results reveal that gluino or $\rho$-chargino predominantly dictate the predictions for BR($t\rightarrow qV,qh$), while the impact of neutralino or $\chi$-chargino contributions is insignificant. By incorporating the constraints imposed by the squark mixing parameters from $\bar{B}\rightarrow X_s\gamma$ and $B^0_{d,s}\rightarrow \mu^+\mu^-$, the theoretical predictions for BR($t\rightarrow qg$) can be significantly augmented to approximately $\mathcal O(10^{-5}-10^{-6})$. On the other hand, the values of BR($t\rightarrow q\gamma,qZ,qh$) are predicted to be at least four orders of magnitude below the current experimental limits.
\end{abstract}



\maketitle

\section{Introduction\label{sec1}}
The observation of flavor-changing neutral current (FCNC) processes, specifically $t\rightarrow qV,qh$, in experiments would provide significant evidence for the existence of new physics. In the context of the standard model (SM), these FCNC processes are prohibited at the tree level and heavily suppressed at the one-loop level by the GIM mechanism. The branching ratios of the FCNC decays $t\rightarrow qV,qh$ in the SM \cite{Aguilar2004,Balaji2020}, as well as their corresponding experimental bounds \cite{PDG2022}, are summarized in TABLE.\ref{tab1}. It is evident that the SM predictions for these branching ratios are considerably lower than the current experimental bounds.
\begin{table*}[h]
\caption{The SM predictions and experimental bounds on the FCNC decays $t\rightarrow qV,qh$.}
\label{tab1}
\tabcolsep 5pt 
\begin{tabular*}{\textwidth}{cccccccc}
\hline
Decay&SM&Bound&Experiment&Decay&SM&Bound&Experiment\\\hline
$t\rightarrow u\gamma$&$3.7\times10^{-16}$&$6.1\times 10^{-5}$&ATLAS \cite{Aad2020}&$t\rightarrow c\gamma$&$4.6\times10^{-14}$&$2.2\times 10^{-4}$&ATLAS \cite{Aad2020}\\
$t\rightarrow ug$&$3.7\times10^{-14}$&$2\times 10^{-5}$&CMS \cite{Khachatryan2017}&$t\rightarrow cg$&$4.6\times10^{-12}$&$4.1\times 10^{-5}$&CMS \cite{Khachatryan2017}\\
$t\rightarrow uZ$&$8\times10^{-17}$&$1.5\times 10^{-4}$&CMS \cite{CMS017}&$t\rightarrow cZ$&$1\times10^{-14}$&$3.7\times 10^{-4}$&CMS \cite{CMS017}\\
$t\rightarrow uh$&$2\times10^{-17}$&$1.9\times 10^{-4}$&CMS \cite{CMS2022}&$t\rightarrow ch$&$3\times10^{-15}$&$7.3\times 10^{-4}$&CMS \cite{CMS2022}\\
\hline
\end{tabular*}
\end{table*}
Searches for top quark FCNC decays are an integral part of the experimental program at the High-Luminosity Large Hadron Collider (HL-LHC), High Energy Large Hadron Collider (HE-LHC), and the proposed Future Circular Collider in Hadron-Hadron mode (FCC-hh). TABLE.\ref{tabfs} gives an overview of the future experimental sensitivities for the FCNC decays $t\rightarrow qV,qh$. By comparing the TABLE.I and TABLE.II, we can find the projected values of the branching ratios are about one or two orders of magnitude below the current experimental bounds.
\begin{table}[h]
\caption{The future experimental sensitivities for the FCNC decays $t\rightarrow qV,qh$.}
\label{tabfs}
\tabcolsep 9pt 
\begin{tabular*}{1\textwidth}{cccccc}
\hline
Decay&Bound&Experiment&Decay&Bound&Experiment\\\hline
$t\rightarrow u\gamma$&$8.6\times 10^{-6}$&HL-LHC \cite{CERN2019}&$t\rightarrow c\gamma$&$7.4\times 10^{-5}$&HL-LHC \cite{CERN2019}\\
&$9.1\times 10^{-7}$& FCC-hh \cite{Aguilar2017}&&$2.3\times 10^{-5}$& FCC-hh \cite{Aguilar2017}\\
$t\rightarrow ug$&$3.8\times 10^{-6}$&HL-LHC \cite{CMSqg}&$t\rightarrow cg$&$3.2\times 10^{-5}$&HL-LHC \cite{CMSqg}\\
&$5.18\times 10^{-7}$&FCC-hh \cite{Oyulmaz}&&$4.45\times 10^{-7}$&FCC-hh \cite{Oyulmaz}\\
$t\rightarrow uZ$&$(2.4-5.8)\times10^{-5}$&HL-LHC \cite{CERN2019}&$t\rightarrow cZ$&$(2.4-5.8)\times 10^{-5}$&HL-LHC \cite{CERN2019}\\
&$2.7\times10^{-6}$&FCC-hh \cite{Aguilar2017}&&$5.0\times 10^{-5}$&FCC-hh \cite{Aguilar2017}\\
$t\rightarrow uh$&$4.4\times10^{-5}$&HE-LHC\cite{YJZhang}&$t\rightarrow ch$&$6.4\times 10^{-5}$&HE-LHC\cite{YJZhang}\\
&$1.3\times10^{-5}$&FCC-hh\cite{YJZhang}&&$1.6\times 10^{-5}$&FCC-hh\cite{YJZhang}\\
\hline
\end{tabular*}
\end{table}

In the literature, the branching ratios of $t\rightarrow qV,qh$ induced by FCNC interactions can be enhanced close to the experimental limits in various scenarios beyond the SM. Several models have been proposed to explain these enhancements. These include two-higgs doublet models \cite{Eilam1991, Diaz1990, Grzadkowski1991, Bejar2001, Han2014, Cai2022}, Minimal Supersymmetric Standard Model (MSSM) \cite{Li1994, Yang1995, Guasch1999, Yang1998, Cao2006, Cao2007, Eilam2001, Couture1995, Couture1997, Lopez1997, Liu2004, Delepine2004, Dedes2014}, littlest higgs model with T-parity \cite{Hou2007, Yang2009, Yang2014}, left-right supersymmetric model \cite{Frank2005}, topcolor-assisted technicolor model \cite{Lu2003}, MSSM with local $U(1)_{B-L}$ gauge symmetry \cite{Yang2018}, extra dimensional models \cite{Agashe, Gao, Chiang}, leptoquark model \cite{Bolanos} and dark matter model \cite{liu2022}.
In supersymmetric models, the FCNC decays $t\rightarrow qV,qh$ can arise from various diagrams involving neutralinos, charginos, and gluinos. Both the supersymmetric electroweak sector and the supersymmetric QCD sector can contribute significantly to the corrections in these decays within the MSSM \cite{Li1994, Lopez1997, Yang1995, Guasch1999}. The effects of left-handed and right-handed squark mixing on BR($t\rightarrow cV$) have been studied in \cite{Couture1995, Couture1997}. In the general unconstrained MSSM, where the soft SUSY breaking parameters allow for flavor-dependent mixings in the squark mass matrix, the predicted BR($t\rightarrow cV$) strongly depend on the soft trilinear couplings $A_U$ \cite{Liu2004}. The effects of holomorphic and non-holomorphic trilinear soft SUSY breaking terms on BR($t\rightarrow qh$) have been discussed in \cite{Dedes2014}. In the MSSM with R-parity violation, $t\rightarrow qV,qh$ can also proceed through R-parity violating interactions \cite{Yang1998, Eilam2001}.

In this paper, we present an analysis of the FCNC decays $t\rightarrow qV,qh$ in the context of the Minimal R-symmetric Supersymmetric Standard Model (MRSSM) \cite{Kribs}. The MRSSM is designed to address the flavor problem in the MSSM by incorporating an unbroken global $U(1)_R$ symmetry. This R-symmetry forbids certain terms in the model, such as Majorana gaugino masses, $\mu$ term, $A$ terms, and left-right squark and slepton mass mixings. Instead, Dirac mass terms are introduced to generate masses for neutralinos and gluinos. Notably, the neutralinos and gluinos in the MRSSM are of the Dirac type, with the particle and its antiparticle differing by a factor of $-$1 in R-charge.
The absence of certain soft breaking trilinear terms that contribute to flavor-violating observables in the MSSM alleviates the flavor problem in the MRSSM \cite{Kribs}. Moreover, the MRSSM features an enlarged number of chargino degrees of freedom, which are categorized into $\chi$-charginos and $\rho$-charginos based on their R-charge. While $\chi$-charginos can contribute to lepton and quark flavor-violating processes, $\rho$-charginos primarily impact quark flavor-violating phenomena. Consequently, the MRSSM exhibits distinct phenomenology compared to the MSSM. Various studies exploring the phenomenology of the MRSSM can be found in the literature \cite{Diessner2014, Diessner2015, Diessner2016, Diessner2017, Diessner2019, Diessner20192, Kotlarski, Kumar, Blechman, Kribs1, Frugiuele, Jan, Chakraborty, Braathen, Athron2017, Athron2022,Alvarado,sks1,sks2,sks3}.

In our analysis, we calculate the branching ratios for the FCNC decays $t\rightarrow qV,qh$ in the MRSSM using an effective Lagrangian approach. We consider contributions from various particles, including squarks, gluinos, $\chi$-charginos, $\rho$-charginos, and neutralinos, at the one-loop level.
To obtain realistic estimates for the BR($t\rightarrow qV,qh$), we take into account experimental constraints on the parameter space. These constraints arise from observables such as the Higgs mass, W boson mass, and low-energy B meson physics processes like $\bar{B}\rightarrow X_s\gamma$ and $B^0_s\rightarrow \mu^+\mu^-$. By considering these constraints, we ensure that the parameter values used in our calculations are consistent with experimental data.
In our study, we investigate the dependence of the BR($t\rightarrow qV,qh$) on several key parameters, namely, the ratio tan$\beta$, the third-generation squark mass $m_{\tilde{b}}$, the gluino mass $m_{\tilde{g}}$, and the off-diagonal parameter $\delta$ in the squark mass matrix. By varying these parameters, we explore how the branching ratios change and identify the dominant contributions to the FCNC decays $t\rightarrow qV,qh$ in the MRSSM.
Through our analysis, we aim to provide valuable insights into the behavior and characteristics of the BR($t\rightarrow qV,qh$) within the context of the MRSSM.

The outline of this article is as follows. In Section \ref{sec2}, we present the details of the MRSSM. All relevant mass matrices and mixing matrices are provided. The Feynman diagrams contributing to $t\rightarrow qV,qh$ in the MRSSM are given at one loop level. Notations and conventions for effective operators and Wilson coefficients are listed. The numerical results are presented in Section \ref{sec3}. The conclusion is drawn in Section \ref{sec4}.

\section{MRSSM\label{sec2}}

In this section, we provide a concise overview of the MRSSM and clarify the notation that will be used throughout the rest of the work. The MRSSM is an extension of the SM and the MSSM, sharing the same gauge group $SU(3)_C\times SU(2)_L\times U(1)_Y$.
In addition to the matter fields present in the MSSM, the MRSSM incorporates extra Higgs and gauge superfields introduced through chiral adjoints $\hat{\cal O},\hat{T},\hat{S}$ and two $R$-Higgs iso-doublets $\hat{R}_{u}$ and $\hat{R}_{d}$. A comprehensive list of superfields and component fields in the MRSSM can be found in TABLE.\ref{tab2}. All the SM fields carry an R-charge of 0. The scalar component field has the same R-charge as its superfield. The fermionic components R-charge is reduced by 1 and the auxiliary fields R-charge by 2.
\begin{table}[h]
\footnotesize
\caption{The R-charges of the superfields and the corresponding bosonic and fermionic components in the MRSSM.}
\label{tab2}
\doublerulesep 0.1pt
\tabcolsep 5pt 
\begin{tabular}{cccccccc}
\hline
Field&Superfield&R-charge&Boson&R-charge&Fermion&R-charge\\
\hline
Gauge sector & $\hat{g},\hat{W},\hat{B}$&0& $g,W,B$&0& $\tilde{g},\tilde{W}\tilde{B}$&  $+$1 \\\hline
\multirow{2}*{Matter}& $\hat{l}, \hat{e}$& $+$1& $\tilde{l},\tilde{e}^*_R$&$+$1& $l,e^*_R$& 0   \\
 \cline{2-7}     & $\hat{q},{\hat{d}},{\hat{u}}$& $+$1& $\tilde{q},{\tilde{d}}^*_R,{\tilde{u}}^*_R$ & $+$1& $q,d^*_R,u^*_R$                             & 0 \\\hline
H-Higgs & ${\hat{H}}_{d,u}$&0& $H_{d,u}$& 0& ${\tilde{H}}_{d,u}$&$-$1 \\ \hline
R-Higgs & ${\hat{R}}_{d,u}$ & $+$2& $R_{d,u}$&$+$2& ${\tilde{R}}_{d,u}$& $+$1 \\\hline
Adjoint chiral& $\hat{\cal O},\hat{T},\hat{S}$&0& $O,T,S$&0& $\tilde{O},\tilde{T},\tilde{S}$ &$-$1 \\
\hline
\end{tabular}
\end{table}

The superpotential of the MRSSM, as described in \cite{Diessner2014}, is given by the following expression
\begin{align}
\mathcal{W}_{MRSSM} = &\Lambda_d(\hat{R}_d\hat{T})\hat{H}_d+\Lambda_u(\hat{R}_u\hat{T})\hat{H}_u\nonumber\\
&+\mu_d(\hat{R}_d\hat{H}_d)+\mu_u(\hat{R}_u\hat{H}_u)\nonumber\\
&+\lambda_d\hat{S}(\hat{R}_d\hat{H}_d)+\lambda_u\hat{S}(\hat{R}_u\hat{H}_u)\nonumber\\
&-Y_d\hat{d}(\hat{q}\hat{H}_d)-Y_e\hat{e}(\hat{l}\hat{H}_d)+Y_u\hat{u}(\hat{q}\hat{H}_u),
\label{sptl}
\end{align}
where $\hat{H}_u$ and $\hat{H}_d$ represent the MSSM-like Higgs weak iso-doublets, $\hat{R}_u$ and $\hat{R}_d$ are the R-charged Higgs $SU(2)_L$ doublets. These R-charged Higgs doublets are responsible for generating the Dirac higgsino mass parameters $\mu_u$ and $\mu_d$. Although R-symmetry forbids the $\mu$ terms of the MSSM, the bilinear combinations of the normal Higgs $SU(2)_L$ doublets $\hat{H}_u$ and $\hat{H}_d$ with the Higgs $SU(2)_L$ doublets $\hat{R}_u$ and $\hat{R}_d$ are allowed in Eq.(\ref{sptl}). The trilinear couplings $\lambda_u$, $\lambda_d$, $\Lambda_u$ and $\Lambda_d$ are Yukawa-like terms involving the singlet $\hat{S}$ and the triplet $\hat{T}$, and play a significant role in obtaining a 125 GeV Higgs boson mass.

The soft-breaking terms involving scalar mass are \cite{Diessner2014}
\begin{align}
V_{SB,S} = &m^2_{H_d}(|H^0_d|^2+|H^{-}_d|^2)+m^2_{H_u}(|H^0_u|^2+|H^{+}_u|^2)\nonumber\\
&+(B_{\mu}(H^-_dH^+_u-H^0_dH^0_u)+h.c.)\nonumber\\
&+m^2_{R_d}(|R^0_d|^2+|R^{+}_d|^2)+m^2_{R_u}(|R^0_u|^2+|R^{-}_u|^2)\nonumber\\
&+m^2_T(|T^0|^2+|T^-|^2+|T^+|^2)+m^2_S|S|^2\nonumber\\
&+ m^2_O|O^2|+\tilde{d}^*_{L,i} m_{q,{i j}}^{2} \tilde{d}_{L,j} +\tilde{d}^*_{R,i} m_{d,{i j}}^{2} \tilde{d}_{R,j}\nonumber\\
&+\tilde{u}^*_{L,i}  m_{q,{i j}}^{2} \tilde{u}_{L,j}+\tilde{u}^*_{R,i}  m_{u,{i j}}^{2} \tilde{u}_{R,j}+\tilde{e}^*_{L,i} m_{l,{i j}}^{2} \tilde{e}_{L,j}\nonumber\\
&+\tilde{e}^*_{R,{i}} m_{r,{i j}}^{2} \tilde{e}_{R,{j}} +\tilde{\nu}^*_{L,i} m_{l,{i j}}^{2} \tilde{\nu}_{L,j}.
\label{soft}
\end{align}
In Eq.(\ref{soft}), it can be observed that all trilinear scalar couplings involving Higgs bosons to squarks and sleptons are forbidden. This is because the sfermions possess an R-charge, making these terms non-R-invariant. The Dirac nature is a manifest feature of the MRSSM fermions and the soft-breaking Dirac mass terms of the singlet $\hat{S}$, triplet $\hat{T}$ and octet $\hat{O}$ take the form as
\begin{align}
V_{SB,DG} = &M^B_D(\tilde{B}\tilde{S}-\sqrt{2}\mathcal{D}_B S)
+M^W_D(\tilde{W}^a\tilde{T}^a-\sqrt{2}\mathcal{D}_W^a T^a)\nonumber\\
&+M^O_D(\tilde{g}\tilde{O}-\sqrt{2}\mathcal{D}_g^a O^a)+h.c.,
\label{}
\end{align}
where $\tilde{B}$, $\tilde{W}$ and $\tilde{g}$ are usually MSSM Weyl fermions, $M^B_D$, $M^W_D$ and $M^O_D$ are the bino mass, the wino mass and the gluino mass, respctively. It is important to note that the R-Higgs bosons do not acquire vacuum expectation values, as they carry an R-charge of 2. However, after electroweak symmetry breaking, the vacuum expectation values of the singlet and triplet fields effectively modify the Higgsino mass parameters $\mu_u$ and $\mu_d$. The modified parameters are given by a relationship that can be expressed as follows:
\begin{align}
\begin{array}{l}
\mu_d^{eff,+}= \frac{1}{2} \Lambda_d v_T  + \frac{1}{\sqrt{2}} \lambda_d v_S  + \mu_d,\\[4pt]
\mu_u^{eff,-}= -\frac{1}{2} \Lambda_u v_T  + \frac{1}{\sqrt{2}} \lambda_u v_S  + \mu_u.
\end{array}
\label{}
\end{align}
The quantities $v_T$ and $v_S$ represent the vacuum expectation values of the fields $\hat{T}$ and $\hat{S}$, respectively. These fields have an R-charge of zero.

In the MRSSM, the number of neutralino degrees of freedom is doubled compared to the MSSM due to the Dirac nature of the neutralinos. The neutralinos are composed of four neutral electroweak two-component fermions in the weak basis.
The weak basis consists of $\xi_i$ = ($\tilde{B}$, $\tilde{W}^0$, $\tilde{R}^0_d$, $\tilde{R}^0_u$), which have an R-charge of 1, and $\varsigma_i$ = ($\tilde{S}$, $\tilde{T}^0$, $\tilde{H}^0_d$, $\tilde{H}^0_u$), which have an R-charge of -1.
The neutralino mass matrix and the corresponding diagonalization procedure can be described as
\begin{align}
m_{\chi}=
\left(
\begin{array}{cccc}
M^{B}_D &0 &-\frac{1}{2} g_1 v_d  &\frac{1}{2} g_1 v_u \\
0 &M^{W}_D &\frac{1}{2} g_2 v_d  &-\frac{1}{2} g_2 v_u \\
- \frac{1}{\sqrt{2}} \lambda_d v_d  &-\frac{1}{2} \Lambda_d v_d  &-\mu_d^{eff,+}&0\\
\frac{1}{\sqrt{2}} \lambda_u v_u  &-\frac{1}{2} \Lambda_u v_u  &0 &\mu_u^{eff,-}\end{array}
\right)
\end{align}
and
\begin{align}
(N^{1})^{\ast} m_{\chi} (N^{2})^{\dagger} =m_{\chi}^{\textup{diag}}.
\end{align}
The mass eigenstates $\kappa_i$ and $\varphi_i$, and physical four-component Dirac neutralinos are
\begin{equation}
\xi_i=\sum^4_{j=1}(N^1_{ji})^{\ast}\kappa_j,\;\;\;
\varsigma_i=\sum^4_{j=1}(N^2_{ij})^{\ast}\varphi_j,\;\;\;
\chi^0_i=\left(
\begin{array}{c}
\kappa_i\\
\varphi_i^{\ast}\\
\end{array}
\right).
\end{equation}
The ratio of the Higgs doublet vacuum expectation values is defined by $tan\beta$=$\frac{v_u}{v_d}$.

The number of chargino degrees of freedom in the MRSSM is also doubled compared to the MSSM due to the Dirac nature of the charginos. These charginos can be grouped to two separated chargino sectors based on their R-charge. The $\chi$-chargino sector has an R-charge of 1 electric charge, while the $\rho$-chargino sector has an R-charge of $-$1 electric charge. In the basis $\xi^+_i$=($\tilde{W}^+$, $\tilde{R}^+_d$) and $\varsigma^-_i$=($\tilde{T}^-$, $\tilde{H}^-_d$), the $\chi$-chargino mass matrix and the corresponding diagonalization procedure can be described as
\begin{align}
m_{\chi^{+}} =
\left(
\begin{array}{cc}
g_2 v_T  + M^{W}_D &\frac{1}{\sqrt{2}} \Lambda_d v_d \\
\frac{1}{\sqrt{2}} g_2 v_d  &\mu_d^{eff,-}\end{array}
\right)
\end{align}
and
\begin{align}
(U^{1})^{\ast} m_{\chi^{+}} (V^{1})^{\dagger} =m_{\chi^{+}}^{\textup{diag}}.
\end{align}
The mass eigenstates $\lambda^{\pm}_i$ and physical four-component Dirac charginos are
\begin{equation}
\xi^+_i=\sum^2_{j=1}(V^1_{ij})^{\ast}\lambda^+_j,\;\;\; \varsigma^-_i=\sum^2_{j=1}(U^1_{ji})^{\ast}\lambda^-_j,\;\;\;
\chi^{+}_i=\left(
\begin{array}{c}
\lambda^+_i\\
\lambda_i^{-\ast}\\
\end{array}
\right).
\end{equation}

The $\rho$-chargino mass matrix and the corresponding diagonalization procedure can be described as
\begin{align}
m_{\rho^{-}} =
\left(
\begin{array}{cc}
-g_2 v_T  + M^{W}_D &\frac{1}{\sqrt{2}} g_2 v_u \\
-\frac{1}{\sqrt{2}} \Lambda_u v_u   &-\mu_u^{eff,+}\end{array}
\right)
\end{align}
and
\begin{align}
(U^{2})^{\ast} m_{\rho^{-}} (V^{2})^{\dagger} =m_{\rho^{-}}^{\textup{diag}}.
\end{align}

After the electroweak symmetry breaking, the electroweak scalars with no R-charge acquire vacuum expectation value. The real part of the four complex neutral scalar fields, denoted as ($\phi_d$, $\phi_u$, $\phi_S$, $\phi_T$), corresponds to the two MSSM Higgs doublets $H_{d,u}$ and the N = 2 scalar superpartners of the hypercharge and $SU(2)_L$ gauge fields, $S$ and $T$. The mix of $\phi_d$, $\phi_u$, $\phi_S$ and $\phi_T$ will form four physical mass eigenstates. In the weak basis $(\phi_d,\phi_u,\phi_S,\phi_T)$, the scalar Higgs boson mass matrix and the corresponding diagonalization procedure can be described as
\begin{align}
{\cal M}_h =
\left(
\begin{array}{cc}
{\cal M}_{11}&{\cal M}_{21}^T\\
{\cal M}_{21}&{\cal M}_{22}\\
\end{array}
\right)
\end{align}
and
\begin{align}
{\cal M}_h =
Z^h {\cal M}_{h} (Z^{h})^{\dagger}={\cal M}_{h}^{\textup{diag}},
\end{align}
where the submatrices are
\begin{align}
{\cal M}_{11} =&
\left(
\begin{array}{cc}
m_Z^2 cos^2\beta+m_A^2sin^2\beta&-(m_Z^2+m_A^2)sin\beta cos\beta\\
-(m_Z^2+m_A^2)sin\beta cos\beta &m_Z^2sin^2\beta+m_A^2cos^2\beta\\
\end{array}
\right),\\
{\cal M}_{21} =&
\left(
\begin{array}{cc}
v_d(\sqrt{2}\lambda_d\mu_d^{eff,+}-g_1M_B^D)&
v_u(\sqrt{2}\lambda_u\mu_u^{eff,-}+g_1M_B^D) \\
v_d(\Lambda_d\mu_d^{eff,+}+g_2M_W^D)& -
v_u(\Lambda_u\mu_u^{eff,-}+g_2M_W^D) \\
\end{array}
\right),\\
{\cal M}_{22} =&
\left(
\begin{array}{cc}
4 (M_B^D)^2+m_S^2+\frac{\lambda_d^2 v_d^2+\lambda_u^2 v_u^2}{2} \;
& \frac{\lambda_d \Lambda_d v_d^2-\lambda_u \Lambda_u v_u^2}{2 \sqrt{2}} \\
 \frac{\lambda_d \Lambda_d v_d^2-\lambda_u \Lambda_u v_u^2}{2 \sqrt{2}} \;
 & 4 (M_W^D)^2+m_T^2+\frac{\Lambda_d^2 v_d^2+\Lambda_u^2 v_u^2}{4}\\
\end{array}
\right).
\end{align}
The submatrix ${\cal M}_{11}$ in the left-top corner is MSSM-like.

The mass matrix for up squarks and down squarks, and the corresponding diagonalization procedure can be described as
\begin{align}
m^2_{\tilde{u}} =
\left(
\begin{array}{cc}
(m^2_{\tilde{u}})_{LL} &0 \\
0  &(m^2_{\tilde{u}})_{RR}\end{array}
\right),\;\;\;
m^2_{\tilde{d}}  =
\left(
\begin{array}{cc}
(m^2_{\tilde{d}})_{LL} &0 \\
0  &(m^2_{\tilde{d}})_{RR}\end{array}
\right)
\label{sud}
\end{align}
and
\begin{align}
Z^U m^2_{\tilde{u}} (Z^{U})^{\dagger} =m^{2,\textup{diag}}_{\tilde{u}},\;\;\;Z^D m^2_{\tilde{d}} (Z^{D})^{\dagger} =m^{2,\textup{diag}}_{\tilde{d}},
\end{align}
where the submatrices are
\begin{align}
(m^2_{\tilde{u}})_{LL} =&m_{\tilde{q}}^2+ \frac{1}{2} v_{u}^{2} |Y_{u}|^2
+\frac{1}{24}(g_1^2-3g_2^2)(v_{u}^{2}- v_{d}^{2}) \nonumber\\
&+\frac{1}{3}g_1 v_S M_D^B+g_2v_TM_D^W ,\\
(m^2_{\tilde{u}})_{RR} = & m_{\tilde{u}}^2+\frac{1}{2}v_u^2|Y_u|^2+\frac{1}{6}g_1^2( v_{d}^{2}- v_{u}^{2})-\frac{4}{3} g_1v_SM_D^B,\\
(m^2_{\tilde{d}})_{LL} =&m_{\tilde{q}}^2+ \frac{1}{2} v_{d}^{2} |Y_{d}|^2
+\frac{1}{24}(g_1^2+3g_2^2)(v_{u}^{2}- v_{d}^{2}) \nonumber\\
&+\frac{1}{3}g_1 v_S M_D^B-g_2v_TM_D^W ,\\
(m^2_{\tilde{d}})_{RR} =& m_{\tilde{d}}^2+\frac{1}{2}v_d^2|Y_d|^2+\frac{1}{12}g_1^2( v_{u}^{2}- v_{d}^{2})+\frac{2}{3} g_1v_SM_D^B,
\end{align}
From Eq.(\ref{sud}) we can see that the left-right squark mass mixing is absent in the MRSSM, whereas the $A$ terms are present in the MSSM.

The explicit expressions of the Feynman rules between fermions, sfermions and neutralinos/charginos/gluino are given by \cite{Diessner2014,Kotlarski}
\begin{align}
{\cal L}=&i\bar{u}_i(Y^i_dZ^D_{k3+i}U^1_jP_R)\chi^{+}_j\tilde{d}_k
+i\bar{u}_i(\frac{g_3}{\sqrt{2}}Z^{U\ast}_{k3+i}P_L)\tilde{g}\tilde{u}_k\nonumber\\
&+i\bar{u}_i(Y^i_dV^{2\ast}_{j2}Z^{D\ast}_{ki}P_L-g_2U^{2}_{j1}Z^{D\ast}_{ki}P_R)
\rho^{+}_j\tilde{d}_k \nonumber\\
&+i\bar{u}_i(\frac{2\sqrt{2}}{3}g_1N^{1\ast}_{j1}Z^{U\ast}_{k3+i}P_L
-Y^i_uZ^{U\ast}_{k3+i}N^2_{j4}P_R)\chi^0_j\tilde{u}_k\nonumber\\
&-i\bar{u}_i(\frac{\sqrt{2}}{6}(3g_2N^{1\ast}_{j2}+g_1N^{1\ast}_{j1})Z^U_{ki}P_L
\nonumber\\
&+Y^i_uZ^U_{ki}N^2_{j4}P_R)\chi^{0c}_j\tilde{u}_k
-i\bar{u}_i(\frac{g_3}{\sqrt{2}}Z^U_{ki}P_L)\bar{\tilde{g}}\tilde{u}_k,
\label{frule}
\end{align}
where $P_{L/R}$ is defined as $\frac{1}{2}(1\mp\gamma_5)$.

\begin{figure}[tb]
\centering
\includegraphics[width=0.6\columnwidth]{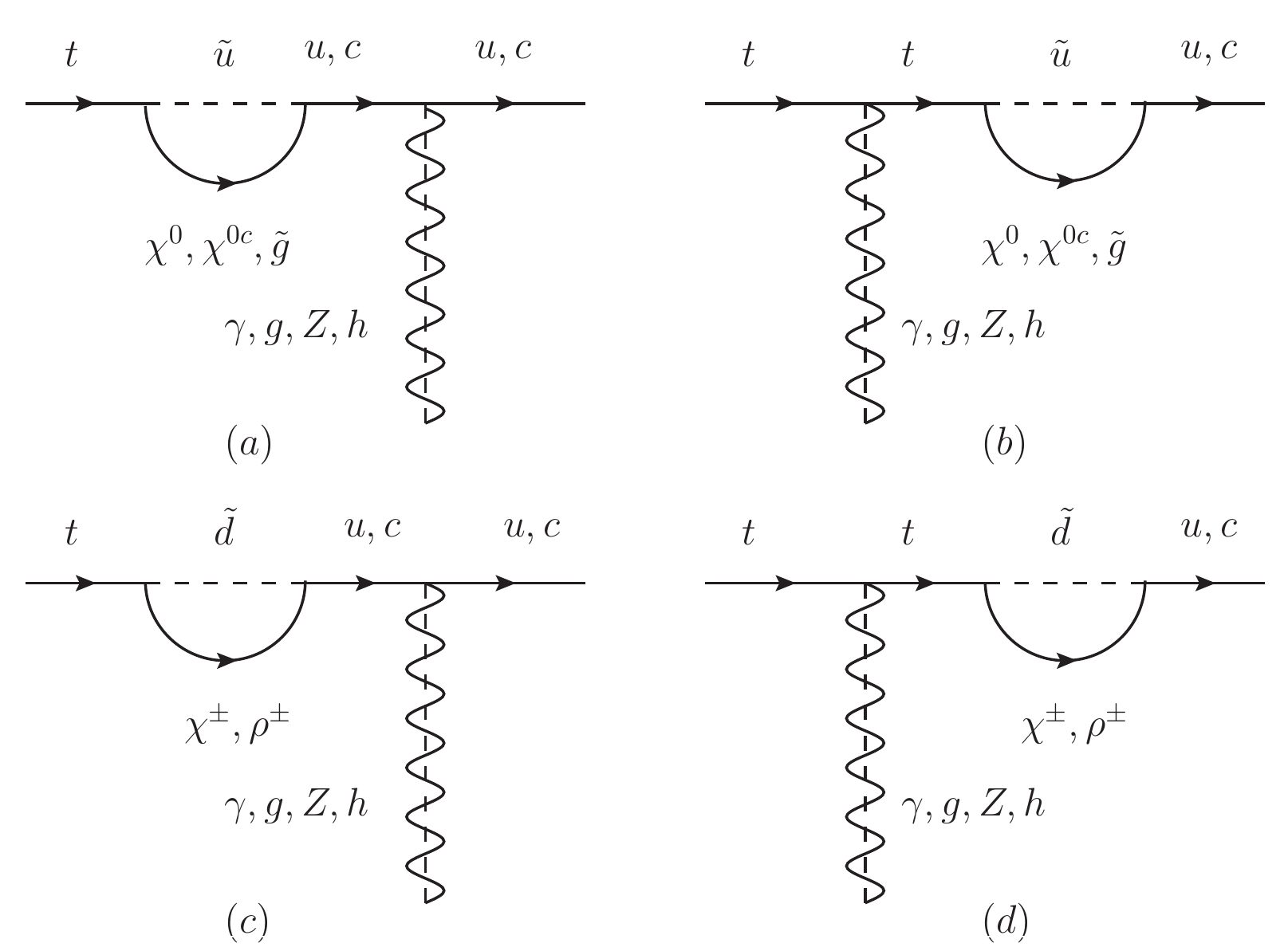}
\caption{The self-energy-type diagrams contributing to $t\rightarrow qV,qh$ in the MRSSM, where $\gamma$, $g$ and $Z$ are denoted by the wavy lines, and $h$ is denoted by the dashed lines.}
\label{Fig1}
\end{figure}
\begin{figure}[tb]
\centering
\includegraphics[width=0.6\columnwidth]{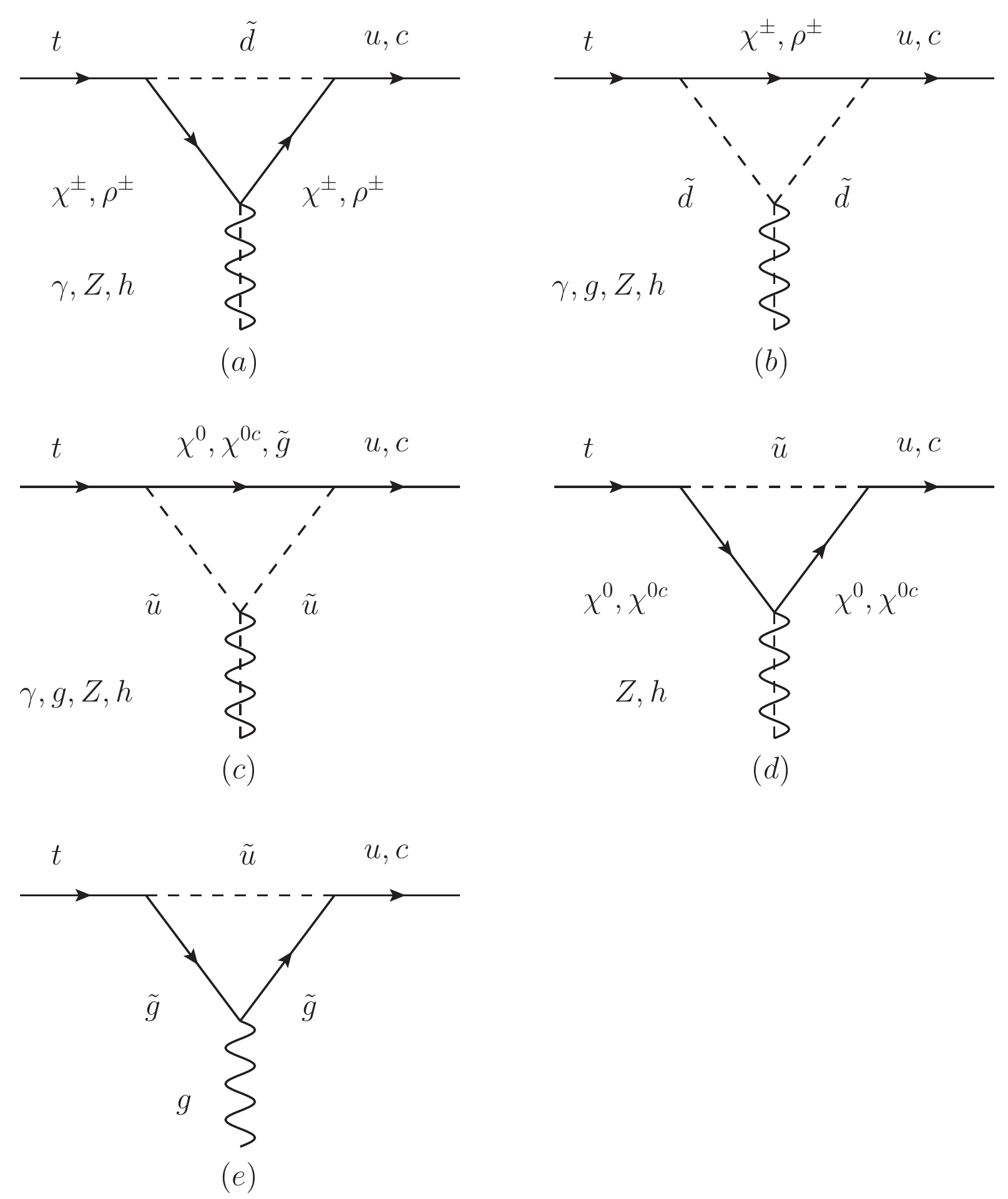}
\caption{The triangle diagrams contributing to $t\rightarrow qV,qh$ in the MRSSM, where $\gamma$, $g$ and $Z$ are denoted by the wavy lines, and $h$ is denoted by the dashed lines.}
\label{Fig2}
\end{figure}

The Feynman diagrams depicting the contributions to $t\rightarrow qV,qh$ in the MRSSM are illustrated in Fig.\ref{Fig1} and Fig.\ref{Fig2}. In these diagrams, $\chi^0$ and $\chi^{0c}$ represent the neutralinos and anti-neutralinos respectively, while $\tilde{g}$ denotes the gluinos and anti-gluinos.
All the self-energy-type diagrams do contribute to the $t\rightarrow qV,qh$ decays. However, not all of the triangle diagrams contribute to these decays due to the absence of coupling between the boson $V/h$ and the loop particles in some decays. For example, referring to Fig.\ref{Fig2}, we observe that the diagram involving two $\tilde{g}$ particles is absent for the $t\rightarrow qZ,qh$ decay. Similarly, the diagrams containing two $\chi^0$/$\chi^{0c}$ or two $\tilde{g}$ particles are not present in the $t\rightarrow q\gamma$ decay, and the diagrams involving two $\chi^{\pm}$, two $\rho^{\pm}$, or two $\chi^0$/$\chi^{0c}$ particles do not contribute to the $t\rightarrow qg$ decay.
It is important to consider these specific variations in the Feynman diagrams when analyzing the $t\rightarrow qV,qh$ decay processes in the MRSSM.

The general form of the amplitude in Fig.\ref{Fig1} and Fig.\ref{Fig2} is given by \cite{Lopez1997,Chiang,Bolanos}
\begin{align}
{\cal M}(t\rightarrow q \gamma) =& \bar{u}_q [ \gamma^{\mu} ( F^L_{1\gamma} P_L + F^R_{1\gamma} P_R )\nonumber\\
&+ i \sigma^{\mu\nu} q_{\nu} ( F^L_{2\gamma} P_L+ F^R_{2\gamma} P_R ) ] u_t \epsilon_{\mu},\label{WCsp}\\
{\cal M}(t\rightarrow q g)=&\bar{u}_q [ T^a \gamma^{\mu} ( F^L_{1g} P_L + F^R_{1g} P_R ) \nonumber\\
&+ i T^a \sigma^{\mu\nu} q_{\nu} ( F^L_{2g} P_L + F^R_{2g} P_R ) ] u_t \epsilon_{\mu},\label{WCsg}\\
{\cal M}(t\rightarrow q Z)=&\bar{u}_q [ \gamma^{\mu} ( F^L_{1 Z} P_L + F^R_{1 Z} P_R ) \nonumber\\
&+ \frac{i \sigma^{\mu\nu} q_{\nu}}{m_t+m_q} ( F^L_{2Z} P_L + F^R_{2Z} P_R ) ] u_t \epsilon_{\mu},\label{WCsZ}\\
{\cal M}(t\rightarrow q h) = & \bar{u}_q ( F^L_{h} P_L + F^R_{h} P_R ) u_t h,
\label{WCsh}
\end{align}
where $q_{\nu}$ is the momentum of the outgoing gauge boson, while $\epsilon_{\mu}$ represents the polarization vector of the outgoing gauge boson. $\sigma^{\mu\nu}$ is given by $\frac{i}{2}[\gamma^{\mu},\gamma^{\nu}]$. $T^a$ denotes the generators of $SU(3)_C$. In the MRSSM, the contribution from the tensor operators in Eq.(\ref{WCsp}) and Eq.(\ref{WCsg}) for $t\rightarrow q\gamma, qg$ can be safely neglected since the coefficients $F^{L/R}_{2\gamma}$ and $F^{L/R}_{2g}$, which are calculated from the Feynman diagrams illustrated in Fig.\ref{Fig1} and Fig.\ref{Fig2}, respectively, are evaluated to be zero.

Consequently, the branching ratios for $t\rightarrow qV,qh$ can be determined as
\begin{align}
BR(t\rightarrow q \gamma) =& \frac{m_t}{32 \pi\Gamma_{t}}[2 (F^L_{1\gamma})^2 + 2 (F^R_{1\gamma})^2+2 m^2_t ( F^L_{2\gamma})^2\nonumber\\
& - 6 m_t (F^L_{1\gamma} F^R_{2\gamma}+ F^R_{1\gamma} F^L_{2\gamma}) +  2 m^2_t ( F^R_{2\gamma})^2],\\
BR(t\rightarrow q g)  =& \frac{m_t}{24 \pi\Gamma_{t}}[2 (F^L_{1g})^2 + 2 (F^R_{1g})^2 +2 m^2_t ( F^L_{2g})^2 \nonumber\\
&- 6 m_t (F^L_{1g} F^R_{2g}+ F^R_{1g} F^L_{2g})+  2 m^2_t ( F^R_{2g})^2],\\
BR(t\rightarrow q Z)  =& \frac{(m_t^2-m^2_Z)^2}{32 \pi m^3_t\Gamma_{t}}[ (2 + \frac{m^2_t}{m^2_Z})(|F^L_{1Z}|^2 + |F^R_{1Z}|^2 )\nonumber\\
&+(2 + \frac{m^2_Z}{m^2_t})(|F^L_{2Z}|^2 + |F^R_{2Z}|^2 )\nonumber\\
&-6Re(F^L_{1Z}F^{R*}_{2Z}+F^R_{1Z}F^{L*}_{2Z})],\\
BR(t\rightarrow q h)  =&\frac{\lambda^{1/2}(m_t^2,m_h^2,m_q^2)}{8\pi m^3_{t}\Gamma_{t}}[2m_tm_q Re(F^L_hF^{R*}_h)\nonumber\\
&+\frac{1}{2}(m_t^2+m_q^2-m_h^2)(|F^L_h|^2+|F^R_h|^2)],
\label{}
\end{align}
where $\Gamma_{t}$ denotes the total width of the top quark. $\lambda(x,y,z)$ is defined as
\begin{align}
\lambda(x,y,z)=x^2+y^2+z^2-2(xy+xz+yz).\nonumber
\end{align}

In the following, we provide the detailed expressions for the Wilson coefficients. The coefficients exhibit left-right symmetry, i.e., $F^{R}_{1V}$ = $F^{L}_{1V}$ ($L \leftrightarrow R$), $F^{R}_{2V}$ = $F^{L}_{2V}$ ($L \leftrightarrow R$), and $F^{R}_{h}$ = $F^{L}_{h}$ ($L \leftrightarrow R$). Any coefficients or charge factors that are not explicitly listed can be assumed to be zero or 1, respectively.

The coefficients corresponding to the Feynman diagrams in Fig.\ref{Fig1}(a) and Fig.\ref{Fig1}(c) are given by
\begin{align}
F^{L,FS}_{1V} =&\frac{Q_{FS}C^{qqV}_R}{m_t^2-m_q^2}(C^{tFS*}_R C^{qFS}_Rm_tm_{F}B_0(m_t^2,m_{F}^2,m_{S}^2)\nonumber\\
&-C^{tFS*}_L C^{qFS}_Rm_t^2B_1(m_t^2,m_{F}^2,m_{S}^2)\nonumber\\
&- C^{tFS*}_R C^{qFS}_L m_t m_{q} B_1(m_t^2,m_{F}^2,m_{S}^2)\nonumber\\
&+C^{tFS*}_L C^{qFS}_L m_{F} m_{q} B_0(m_t^2,m_{F}^2,m_{S}^2)).
\label{cc1}
\end{align}
The coefficients corresponding to the Feynman diagrams in Fig.\ref{Fig1}(b) and Fig.\ref{Fig1}(d) are given by
\begin{align}
F^{L,FS}_{1V} =&\frac{Q_{FS}C^{ttV}_R}{m_q^2-m_t^2}(C^{qFS}_L C^{tFS*}_L m_q m_{F} B_0(m_q^2,m_{F}^2,m_{S}^2)\nonumber\\
&-C^{qFS}_R C^{tFS*}_L m_q^2 B_1(m_q^2,m_{F}^2,m_{S}^2)\nonumber\\
&- C^{qFS}_L C^{tFS*}_R  m_qm_t B_1(m_q^2,m_{F}^2,m_{S}^2)\nonumber\\
&+C^{qFS}_R C^{tFS*}_R m_{F}m_t  B_0(m_q^2,m_{F}^2,m_{S}^2)).
\label{cc2}
\end{align}
The symbols $C^{t...}_{L,R}$ and $C^{q...}_{L,R}$ stand for the Feynman rules between fermions, sfermions and neutralinos/charginos/gluino, and these Feynman rules are given in Eq.(\ref{frule}). The superscript $t$ stands for the top quark. The superscript $q$ stands for the $u$ quark or $c$ quark. In Eq.(\ref{cc1}) and Eq.(\ref{cc2}), $FS$ can take values from the set $\{\chi^{\pm}\tilde{d}$, $\rho^{\pm}\tilde{d}$, $\chi^{0}\tilde{u}$, $\chi^{0c}\tilde{u}$, $\tilde{g}\tilde{u}\}$.
The charge factor $Q_{FS}$ is $4/3$ for $FS\in\{\tilde{g}\tilde{u}\}$ and $F^{L,FS}_{h}$ =$F^{R,FS}_{1V}$$(V\rightarrow h)$.

The coefficients corresponding to the Feynman diagrams in Fig.\ref{Fig2}(b) and Fig.\ref{Fig2}(c) are given by
\begin{align}
F^{L,FS_1S_2}_{1V} =&-2Q_{FS_1S_2}C^{tFS_1*}_LC^{qFS_2}_RC^{S_1S_2V}\nonumber\\
&\times C_{00}(m_{F}^2,m_{S_2}^2,m_{S_1}^2),\\
F^{L,FS_1S_2}_{2Z} =&2Q_{FS_1S_2}C^{qFS_2}_LC^{S_1S_2Z}\nonumber\\
&\times (C^{tFS_1*}_Rm_tC^{\prime}_{0}(m_{F}^2,m_{S_2}^2,m_{S_1}^2)\nonumber\\
&-C^{tFS_1*}_Lm_{F}C^{\prime}_{2}(m_{F}^2,m_{S_2}^2,m_{S_1}^2)),\\
F^{L,FS_1S_2}_h =&-Q_{FS_1S_2}C^{tFS_1*}_LC^{qFS_2}_LC^{S_1S_2h}m_{F_1}\nonumber\\
&\times C_{0}(m_{F}^2,m_{S_2}^2,m_{S_1}^2),
\label{}
\end{align}
where $FS_1S_2$ can take values from the set $\{\chi^{\pm}\tilde{d}\tilde{d}$, $\rho^{\pm}\tilde{d}\tilde{d}$, $\chi^{0}\tilde{u}\tilde{u}$, $\chi^{0c}\tilde{u}\tilde{u}$, $\tilde{g}\tilde{u}\tilde{u}\}$, the charge factor $Q_{FS_1S_2}$ depends on the specific decay channel. For $FS_1S_2 = \tilde{g}\tilde{u}\tilde{u}$, the charge factor $Q_{FS_1S_2}$ is $-1/6$ for $t\rightarrow qg$ and $4/3$ for $t\rightarrow q\gamma$, $qZ$, $qh$.

The coefficients corresponding to the Feynman diagrams in Fig.\ref{Fig2}(a),Fig.\ref{Fig2}(d) and Fig.\ref{Fig2}(e) are given by
\begin{align}
F^{L,F_1F_2S}_{1V} =&-Q_{F_1F_2S}C^{qF_2S}_R(2C^{tF_1S*}_Rm_t\nonumber\\
&\times(C^{F_1F_2V}_Rm_{F_1}C_1(m^2_{F_2},m^2_{F1},m^2_{S})\nonumber\\
&-C^{F_1F_2V}_Lm_{F_2}(C_0(m^2_{F_2},m^2_{F_1},m^2_{S})\nonumber\\
&+C_1(m^2_{F_2},m^2_{F_1},m^2_{S})))-C^{tF_1S*}_L\nonumber\\
&\times(2C^{F_1F_2V}_Lm_{F_1}m_{F_2}C_0(m^2_{F_2},m^2_{F_1},m^2_{S})\nonumber\\
&-C^{F_1F_2V}_R(B_0(0,m^2_{F_1},m^2_{F_2})\nonumber\\
&-2C_{00}(m^2_{F_2},m^2_{F_1},m^2_{S})\nonumber\\
&+m^2_tC_1(m^2_{F_2},m^2_{F_1},m^2_{S})\nonumber\\
&+m^2_{S}C_0(m^2_{F_2},m^2_{F_1},m^2_{S})))),\\
F^{L,F_1F_2S}_{2Z} =&2Q_{F_1F_2S}C^{qF_2S}_L\nonumber\\
&\times(C^{tF_1S*}_RC^{F_1F_2Z}_Lm_tC_{12}(m^2_{F_2},m^2_{F_1},m^2_{S})\nonumber\\
&-C^{tF_1S*}_LC^{F_1F_2Z}_Lm_{F_1} C_{1}(m^2_{F_2},m^2_{F_1},m^2_{S})\nonumber\\
&+C^{tF_1S*}_LC^{F_1F_2Z}_Rm_{F_2}C^{\prime}_{0}(m^2_{F_2},m^2_{F_1},m^2_{S})),\\
F^{L,F_1F_2S}_h=&-Q_{F_1F_2S}C^{tF_1S*}_LC^{qF_2S}_L\nonumber\\
&\times(2C^{F_1F_2h}_Lm_{F_1}m_{F_2} C_0(m^2_{F_2},m^2_{F_1},m^2_{S})\nonumber\\
&+C^{F_1F_2h}_R(m^2_S C_0(m^2_{F_2},m^2_{F_1},m^2_{S})\nonumber\\
&+B_0(0,m^2_{F_1},m^2_{F_2}))),
\label{}
\end{align}
where $F_1F_2S$ can take values from the set $\{\chi^{\pm}\chi^{\pm}\tilde{d}$, $\rho^{\pm}\rho^{\pm}\tilde{d}$, $\chi^{0}\chi^{0}\tilde{u}$, $\chi^{0c}\chi^{0c}\tilde{u}$, $\tilde{g}\tilde{g}\tilde{u}\}$.

The loop integrals $B_{\{0,1\}}$ and $C_{\{0,1,2,00,12,22\}}$ are the Passarino-Veltman functions evaluated in the limit of vanishing external momenta. The loop integrals $C^{\prime}_{\{0,2\}}$ are combinations of $C_{\{0,1,}$ $_{2\}}$ and $C_{\{2,12,22\}}$, respectively. The explicit expressions of these integrals can be found in Refs. \cite{SPheno1,SPheno2,Flavor,Flavor2}.

\section{Numerical Analysis\label{sec3}}

The numerical calculations of the one-loop corrections to the decays $t\rightarrow qV,qh$ in the MRSSM are performed using the SPheno-4.0.5 \cite{SPheno1, SPheno2}. The model implementations are generated using the SARAH-4.15.1 \cite{SARAH, SARAH1, SARAH2, Flavor, Flavor2}. The calculations are carried out within a low-scale version of SPheno, where all free parameters are specified at the SUSY scale.
To incorporate the new observables $t\rightarrow qV,qh$ in SPheno, the details can be found in Ref. \cite{Flavor2, Flavor}.
\begin{align}
&\tan\beta=3,B_\mu=500^2,\lambda_d=1.0,\lambda_u=-0.8,\nonumber\\
&\Lambda_d=-1.2,\Lambda_u=-1.1,M_D^B=550,M_D^W=600,\nonumber\\
& M_D^O=1500,v_S=5.9,\mu_d=\mu_u=500,v_T=-0.38,\nonumber\\
&(m^2_l)_{ii}=(m^2_r)_{ii}=1000^2,(i=1,2,3),m_S=2000,\nonumber\\
&(m^2_{\tilde{q}})_{ii}=(m^2_{\tilde{u}})_{ii}=(m^2_{\tilde{d}})_{ii}=2500^2,(i=1,2),\nonumber\\
&(m^2_{\tilde{q}})_{33}=(m^2_{\tilde{u}})_{33}=(m^2_{\tilde{d}})_{33}=1000^2, m_T=3000.
\label{N1}
\end{align}

The computation relies on a set of benchmark points, which are taken from Ref. \cite{Diessner2014} and Ref. \cite{Diessner2019}, and are provided in Eq.(\ref{N1}). The mass parameters in Eq.(\ref{N1}) are specified in GeV or GeV$^2$. The benchmark points in the MRSSM are carefully chosen to ensure that the model can accommodate a Higgs boson with a mass of around 125 GeV, where the lightest Higgs boson is similar to the one in the SM. The consistency of the Higgs sector with existing experimental data is checked using tools such as HiggsBounds and HiggsSignals. The Higgs potential of the MRSSM is also examined using Vevacious to determine if there are any deeper minima in the parameter space that could lead to issues. By calculation the lightest neutralino can have $75\%$ higgsino-component, $24\%$ bino-component and $1\%$ wino-component. The lighter chargino can have $98\%$ higgsino-component and $2\%$ wino-component. Thus both neutralino and chargino are higgsino-like. We may obtain the same result from Eq.(\ref{N1}), where the higgsino mass $\mu_{d,u}$ is smaller than the wino mass $M_D^W$ and the bino mass $M_D^B$.

The predicted mass of the W boson in the MRSSM is consistent with measurements from various experiments, including the combination of the Large Electron-Positron collider and the Fermilab Tevatron collider \cite{CDF2013}, the ATLAS collaboration \cite{ATLAS2018}, and the LHCb Collaboration \cite{LHCb2022}. By adjusting certain parameters such as $m_{SUSY}$, $v_T$, $\Lambda_u$, and $\Lambda_d$, the recent result on the W boson mass from the CDF collaboration \cite{CDF2022} can also be accommodated within the framework of the MRSSM \cite{Diessner2019, Athron2022}. It is important to note that these parameters have a minimal effect on the prediction of BR($t\rightarrow qV,qh$), which consistently fall within a narrow range of values.

Moreover, the low-energy observables related to B meson physics are also found to be consistent with experimental measurements. Experimental observables such as $\bar{B}\rightarrow X_s\gamma$ and $B^0_{d,s}\rightarrow \mu^+\mu^-$ can be used to constrain the parameter space in the MRSSM. The current experimental data for BR($\bar{B}\rightarrow X_s\gamma$), BR($B^0_s\rightarrow \mu^+\mu^-$), and BR($B^0_d\rightarrow \mu^+\mu^-$) are given in Table.\ref{tab3}. At the chosen benchmark points, the predicted values for BR($\bar{B}\rightarrow X_s\gamma$), BR($B^0_s\rightarrow \mu^+\mu^-$), and BR($B^0_d\rightarrow \mu^+\mu^-$) in the MRSSM are
BR($\bar{B}\rightarrow X_s\gamma$) = $3.47\times 10^{-4}$,
BR($B^0_s\rightarrow \mu^+\mu^-$) = $3.16\times 10^{-9}$,
BR($B^0_d\rightarrow \mu^+\mu^-$) = $1.02\times 10^{-10}$.
These predicted values are within the range of experimental measurements obtained so far.
\begin{table}[h]
\footnotesize
\caption{The experimental bounds on the branching ratios of B meson decays.}
\label{tab3}
\doublerulesep 0.1pt
\tabcolsep 4pt 
\begin{tabular}{cccccc}
\hline
Decay&Branching ratio&Decay&Branching ratio&Decay&Branching ratio\\
\hline
$\bar{B}\rightarrow X_s\gamma$ &  $(3.32\pm 0.15)\times 10^{-4}$ &
$B^0_s\rightarrow \mu^+\mu^-$& $(3.0\pm 0.4)\times 10^{-9} $&
$B^0_d\rightarrow \mu^+\mu^-$& $(1.1^{+ 1.4}_{-1.3})\times 10^{-10}$\\
\hline
\end{tabular}
\end{table}

In the following numerical analysis, the values in Eq.(\ref{N1}) will be used as the default. It is worth mentioning that the off-diagonal elements of the squark mass matrices $m^2_{\tilde{q}}$, $m^2_{\tilde{u}}$, $m^2_{\tilde{d}}$, and the slepton mass matrices $m^2_l$, $m^2_r$ in Eq.(\ref{N1}) are assumed to be zero, which implies the absence of flavor mixing in the squark and slepton sectors.

\begin{figure}[tb]
\centering
\includegraphics[width=0.45\columnwidth]{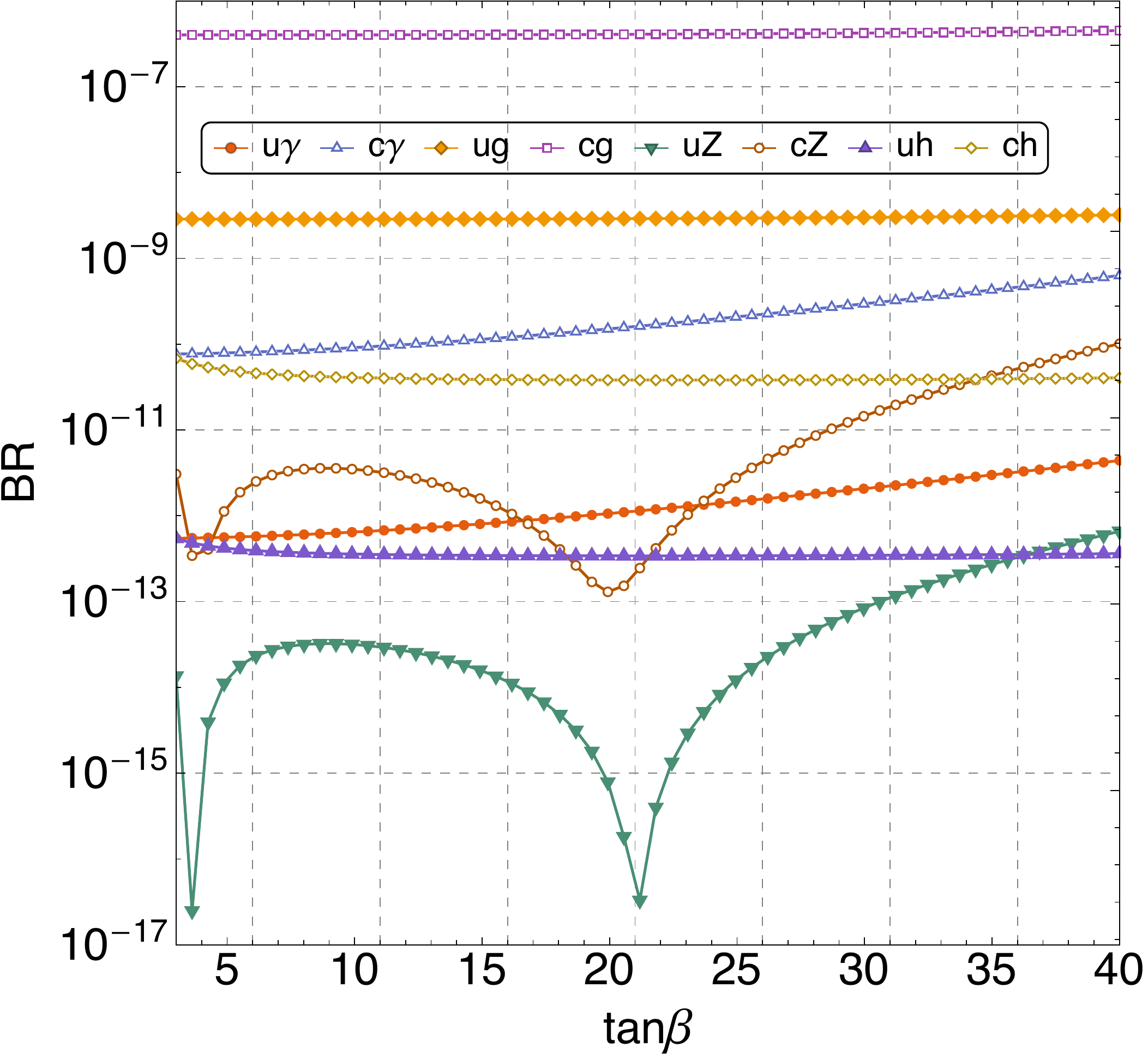}
\caption{The dependence of BR($t\rightarrow qV, qh$) on the ratio tan$\beta$ in the MRSSM. }
\label{figtanb1}
\end{figure}

Using the values provided in Eq.(\ref{N1}), we have plotted the predictions of BR($t\rightarrow qV,qh$) as a function of the ratio tan$\beta$ in Fig.\ref{figtanb1}. It can be observed that the varying ratio tan$\beta$ has negligible impact on the predicted BR($t\rightarrow qg,qh$) values. However, the predicted BR($t\rightarrow q\gamma$) increases gradually by approximately one order of magnitude as tan$\beta$ increases. Interestingly, there are two significant dips in the BR($t\rightarrow qZ$) values at tan$\beta\approx 4$ and tan$\beta\approx 21$.

\begin{figure}[tb]
\centering
\includegraphics[width=0.45\columnwidth]{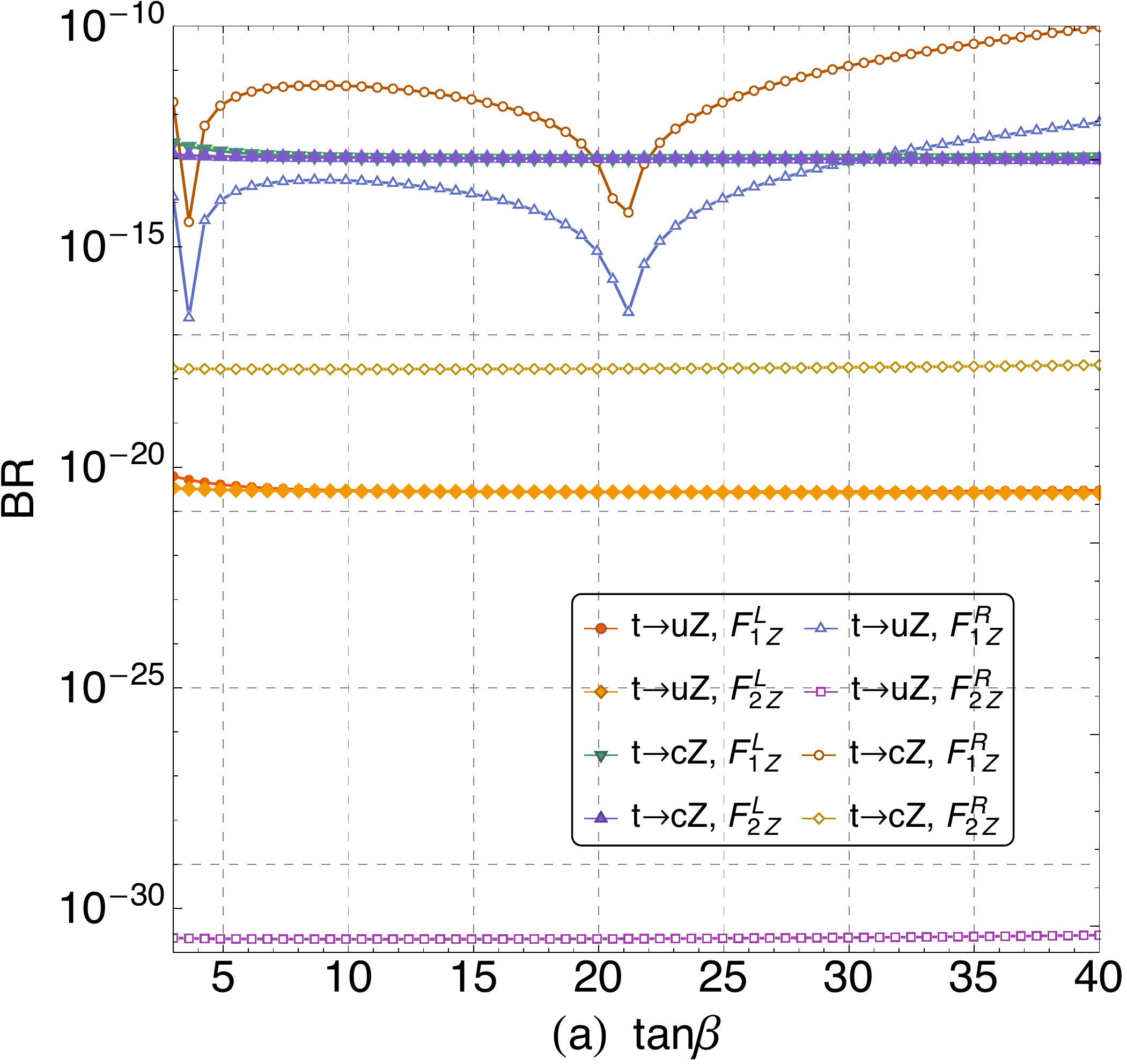}%
\includegraphics[width=0.45\columnwidth]{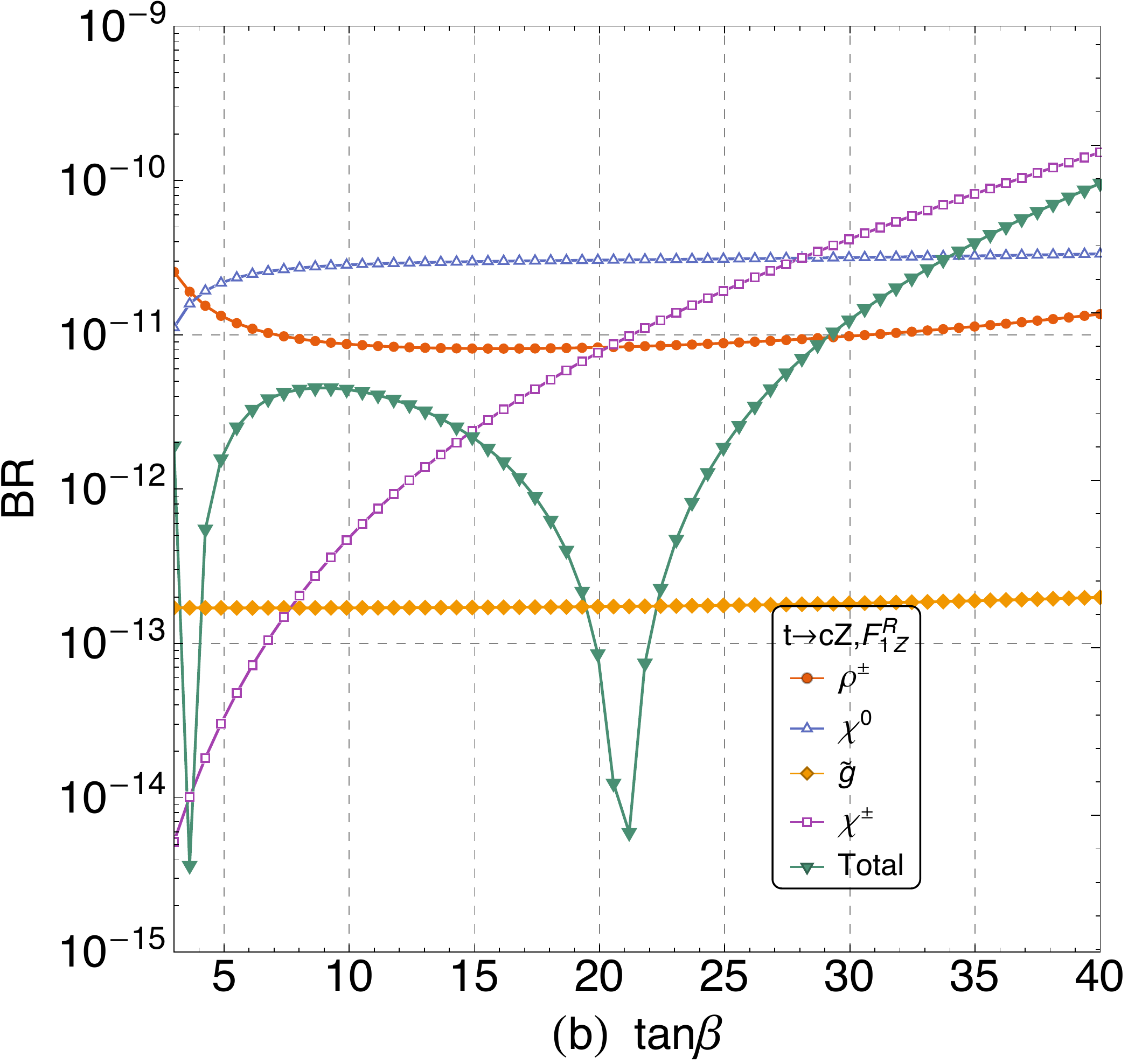}
\caption{The dependence of BR($t\rightarrow qZ, F^{L/R}_{1Z/2Z}$) and BR($t\rightarrow cZ, F^{R}_{1Z}$) on the ratio tan$\beta$ in the MRSSM.}
\label{figtanb2}
\end{figure}

To gain further insights, we consider the contributions from the Wilson coefficients $F^{L/R}_{1Z/2Z}$ separately and plot the predictions of BR($t\rightarrow qZ, F^{L/R}_{1Z/2Z}$) versus tan$\beta$ in Fig.\ref{figtanb2} (a). When considering only one of the coefficients $F^{L}_{1Z}$, $F^{L}_{2Z}$, or $F^{R}_{2Z}$, we find that the varying ratio tan$\beta$ has minimal influence on the predicted BR($t\rightarrow qZ$). However, when $F^{R}_{1Z}$ is included, there are two distinct dips at tan$\beta\approx 4$ and tan$\beta\approx 21$, aligning with the two decreases observed in Fig.\ref{figtanb1}.
In Fig.\ref{figtanb2} (b), we study the predictions of BR($t\rightarrow cZ, F^{R}_{1Z}$) versus tan$\beta$ by considering the contributions from $\rho^{\pm}$, $\chi^0(\chi^{0c})$, $\tilde{g}$, and $\chi^{\pm}$ separately. The coefficients $F^{L}_{1Z}$, $F^{L}_{2Z}$, and $F^{R}_{2Z}$ are set to zero, and it should be noted that the ``Total'' line in Fig.\ref{figtanb2} (b) coincides with the ``$t\rightarrow cZ, F^{R}_{1Z}$'' line in Fig.\ref{figtanb2} (a). As $\chi^{\pm}$-contribution increases with the increase of tan$\beta$, however, $\rho^{\pm}$-contribution, $\chi^0(\chi^{0c})$-contribution and $\tilde{g}$-contribution undergo small changes with the increase of tan$\beta$. The interference among $\chi^{\pm}$, $\rho^{\pm}$, $\chi^{0}$ and gluino would be in disorder. We cannot clearly find the origin of the dips but attribute to the different roles played by $\chi^{\pm}$-contribution at different $tan\beta$. When considering the predicted BR$(t\rightarrow cZ,F^{R}_{1Z})$ with all the contributions, the $\chi^{\pm}$-chargino contribution may play a different role at $\tan\beta\sim $ 4 and $\tan\beta\sim $ 21.

\begin{figure}[tb]
\centering
\includegraphics[width=0.45\columnwidth]{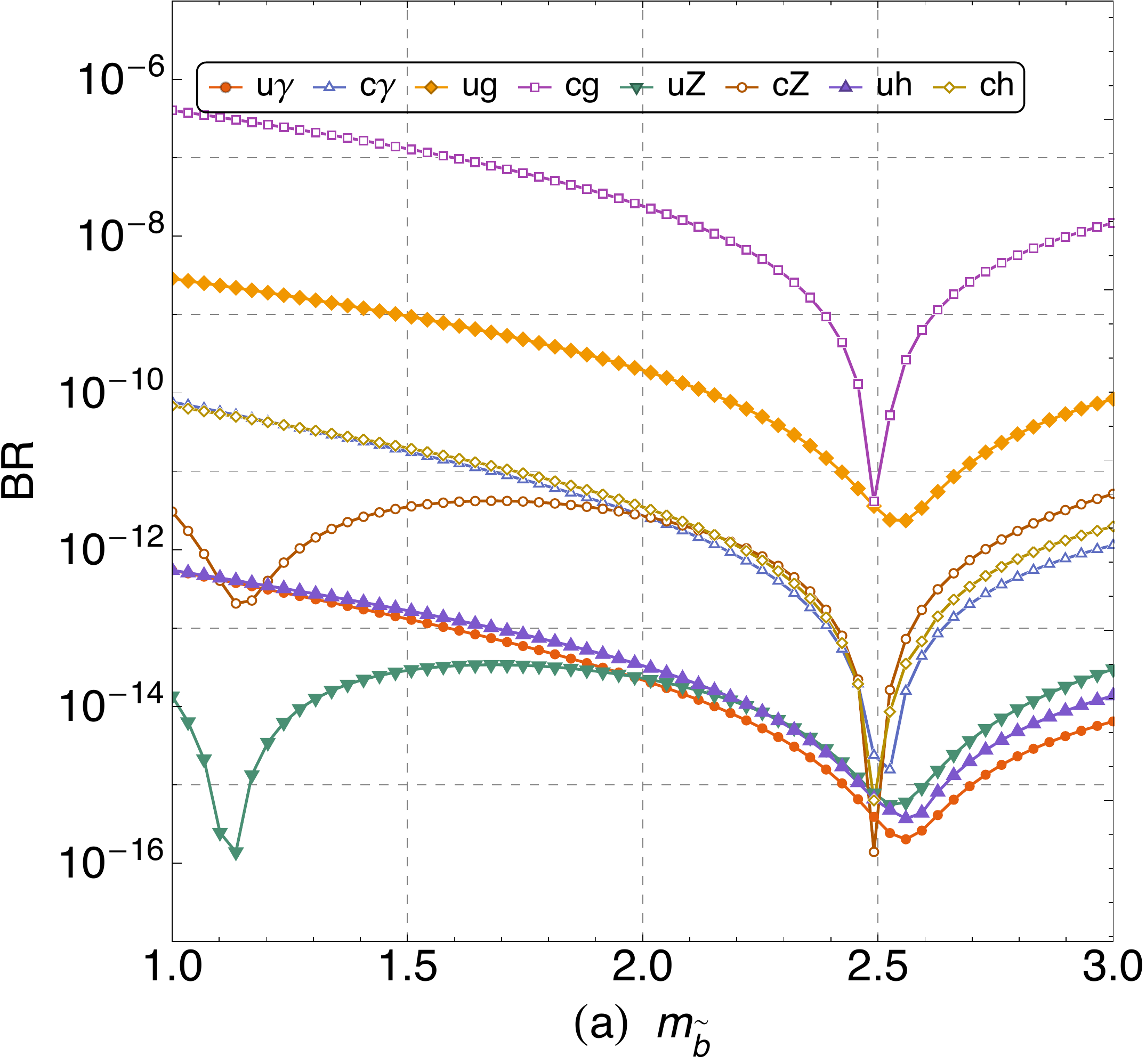}%
\includegraphics[width=0.45\columnwidth]{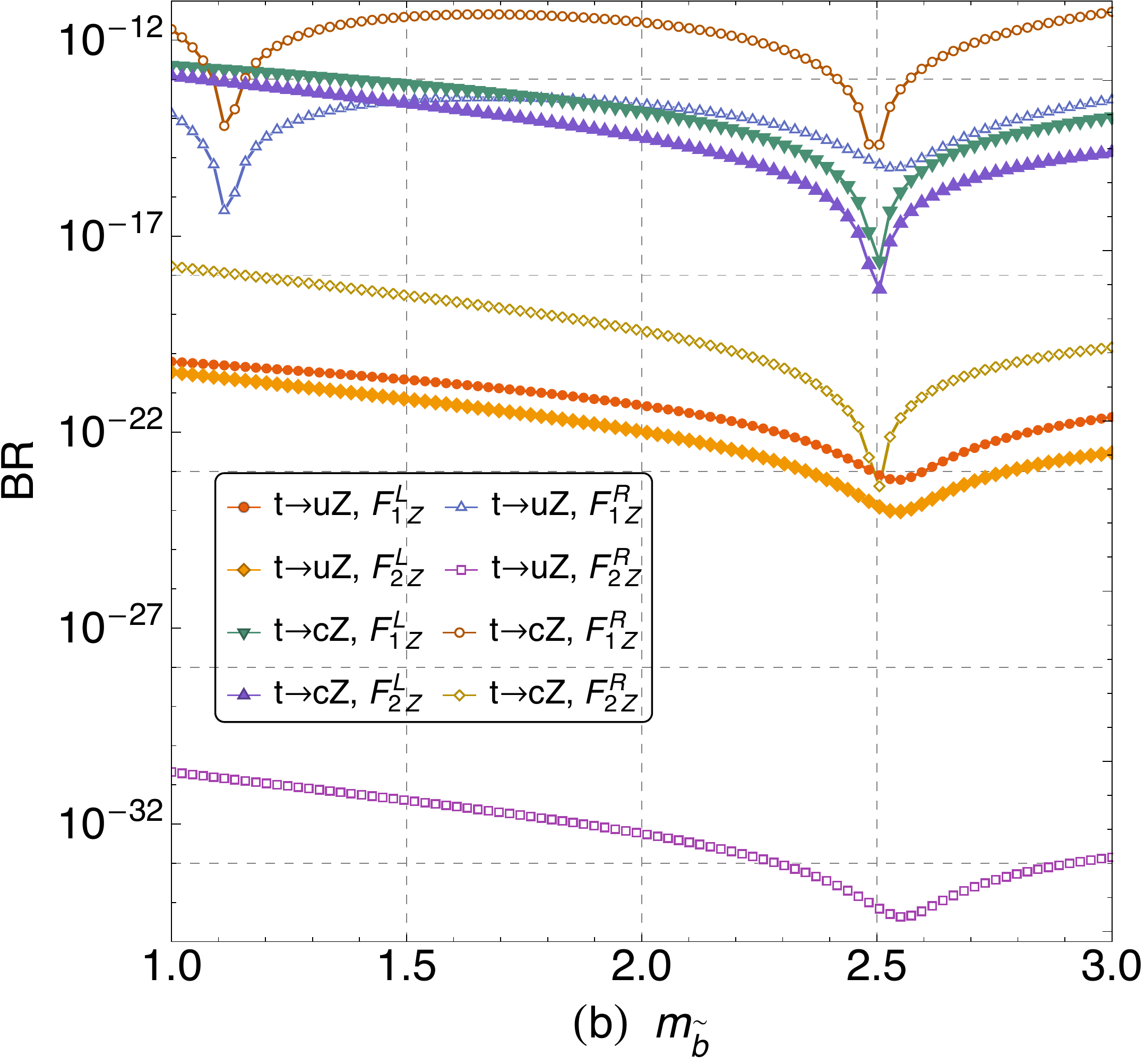}
\caption{The dependence of BR($t\rightarrow qV, qh$) and BR($t\rightarrow qZ, F^{L/R}_{1Z/2Z}$) on the third generation squark mass $m_{\tilde{b}}$ in the MRSSM. The squark mass $m_{\tilde{b}}$ is given in TeV.}
\label{figmb1}
\end{figure}

In Fig.\ref{figmb1} (a), we have plotted the predictions of BR($t\rightarrow qV,qh$) as a function of the third generation squark mass $m_{\tilde{b}}$, where $(m_{\tilde{q}})_{33}$ = $(m_{\tilde{u}})_{33}$ = $(m_{\tilde{d}})_{33}$ = $m_{\tilde{b}}$ are assumed. The mass parameter $m_{\tilde{b}}$ is given in TeV. The other squark masses ($(m_{\tilde{q},\tilde{u},\tilde{d}})_{11}$, $(m_{\tilde{q},\tilde{u},\tilde{d}})_{22}$) are kept at 2.5 TeV as stated in Eq.(\ref{N1}). At $m_{\tilde{b}}$ = 1 TeV, the following hierarchies are observed: BR($t\rightarrow cg$) $>$ BR($t\rightarrow ug$) $>$ BR($t\rightarrow c\gamma$) $>$ BR($t\rightarrow cZ$) $>$ BR($t\rightarrow u\gamma$) $>$ BR($t\rightarrow uZ$) and BR($t\rightarrow ch$) $>$ BR($t\rightarrow uh$). These hierarchies are similar to those found in the SM, as depicted in Table.\ref{tab1}, and some other new physics models \cite{Yang2009,Gao,Chiang, Yang2018}. However, in certain models, these hierarchies may be violated, and the branching ratios BR($t\rightarrow cV, ch$) could be of the same order of magnitude \cite{Lopez1997,Yang1998,Lu2003,Liu2004,Frank2005,Hou2007,Cao2007}. Actually, we have studied the influence of $(m_{\tilde{q}})_{33}$, $(m_{\tilde{u}})_{33}$ and $(m_{\tilde{d}})_{33}$ separately. The result shows $(m_{\tilde{u}})_{33}$ and $(m_{\tilde{d}})_{33}$ have very little effect on the predicted BR($t\rightarrow qV,qh$), which consistently fall within a narrow range of values. The effect of $(m_{\tilde{q}})_{33}$ is same as $m_{\tilde{b}}$.

In Fig.\ref{figmb1} (a), there appears to be a narrow region of cancellation around $m_{\tilde{b}}$ = 2.5 TeV, resulting in predicted BR($t\rightarrow qV,qh$) values that closely align with the SM predictions. Consequently, it may be effectively unobservable in future experiments. This cancellation arises from the squark mass degeneracy present in the matrices $m^2_{\tilde{q}}$, $m^2_{\tilde{d}}$, and $m^2_{\tilde{u}}$. When $m_{\tilde{b}}$ is at 2.5 TeV, the values of $(m_{\tilde{q}})_{33}=(m_{\tilde{u}})_{33}=(m_{\tilde{d}})_{33}$ would be close to $(m_{\tilde{q}})_{ii}=(m_{\tilde{u}})_{ii}=(m_{\tilde{d}})_{ii}$($i$=1,2). This will lead to the mass degeneration of the three generation squark mass and gives rise to the dip at $m_{\tilde{b}}\simeq$ 2.5 TeV.
\begin{figure}[tb]
\centering
\includegraphics[width=2.5in]{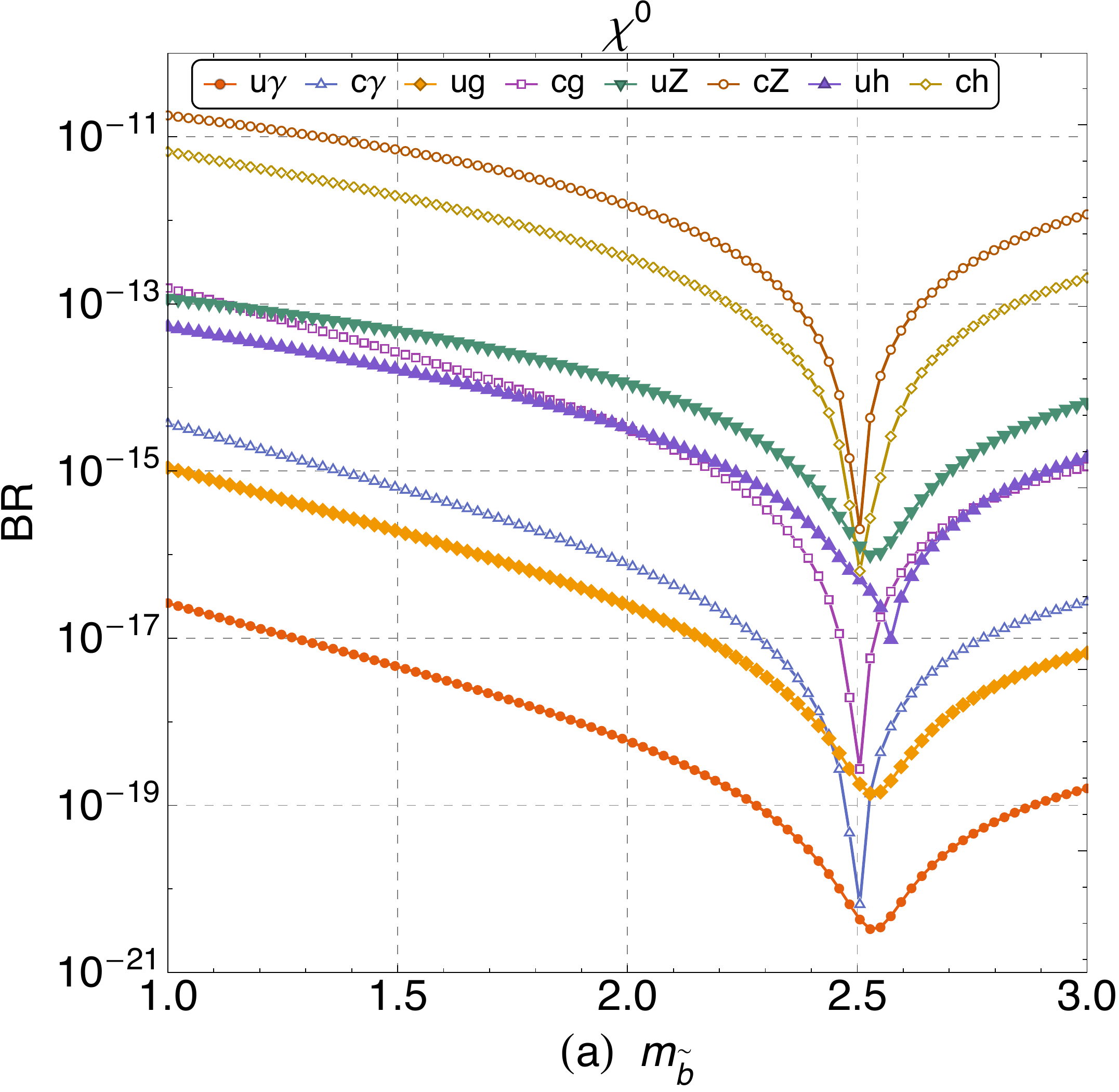}%
\includegraphics[width=2.5in]{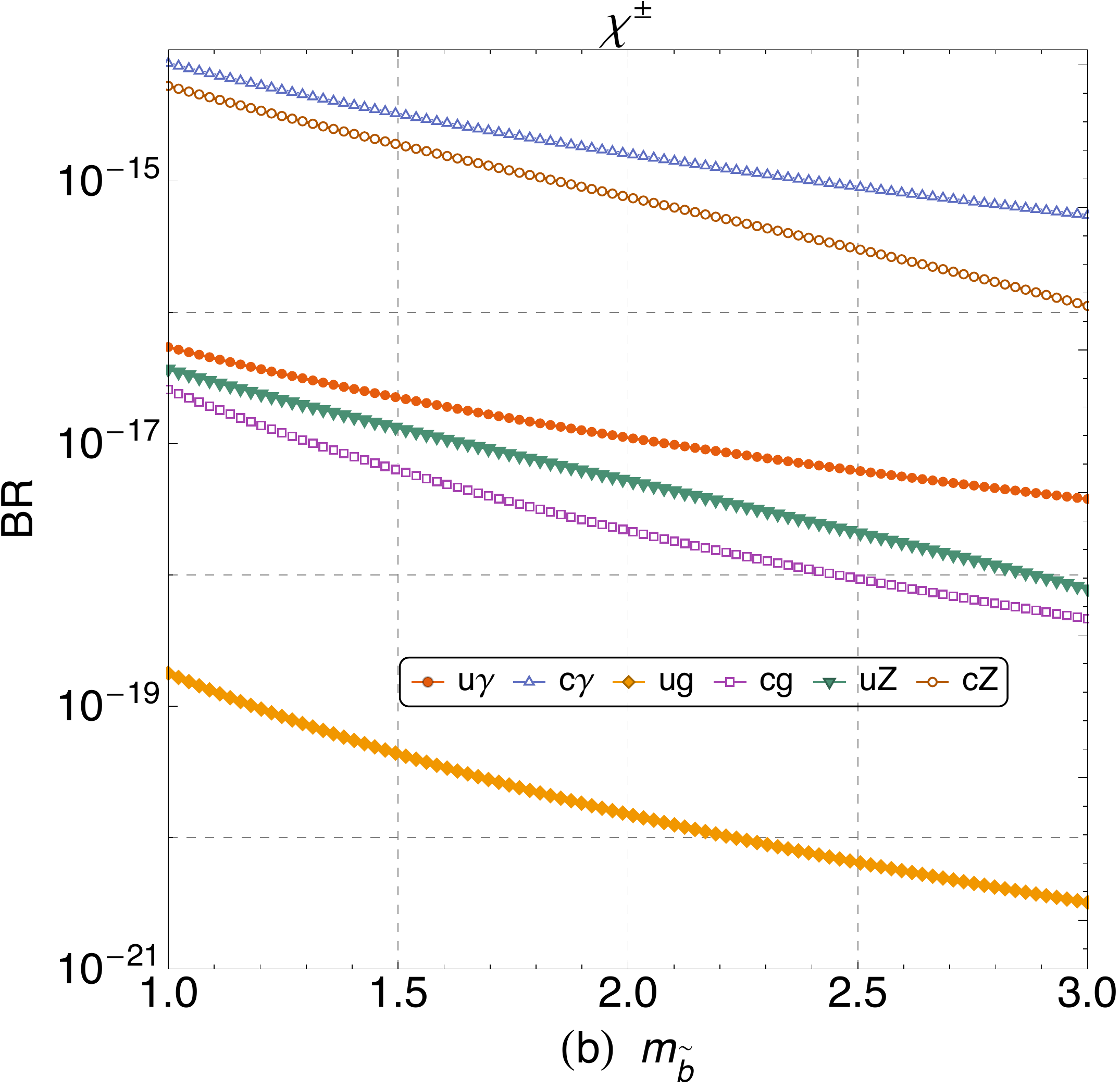}
\includegraphics[width=2.5in]{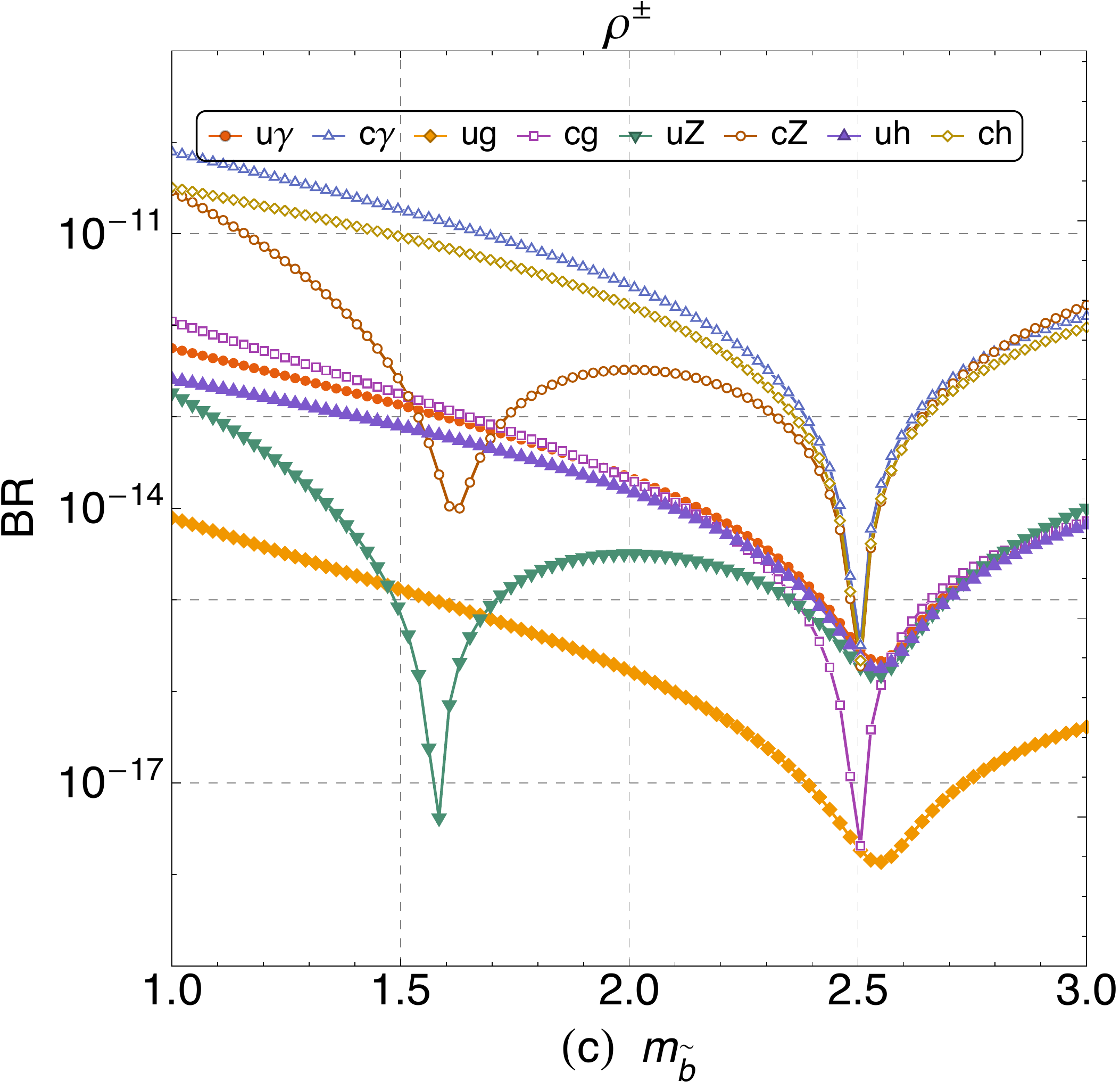}%
\includegraphics[width=2.5in]{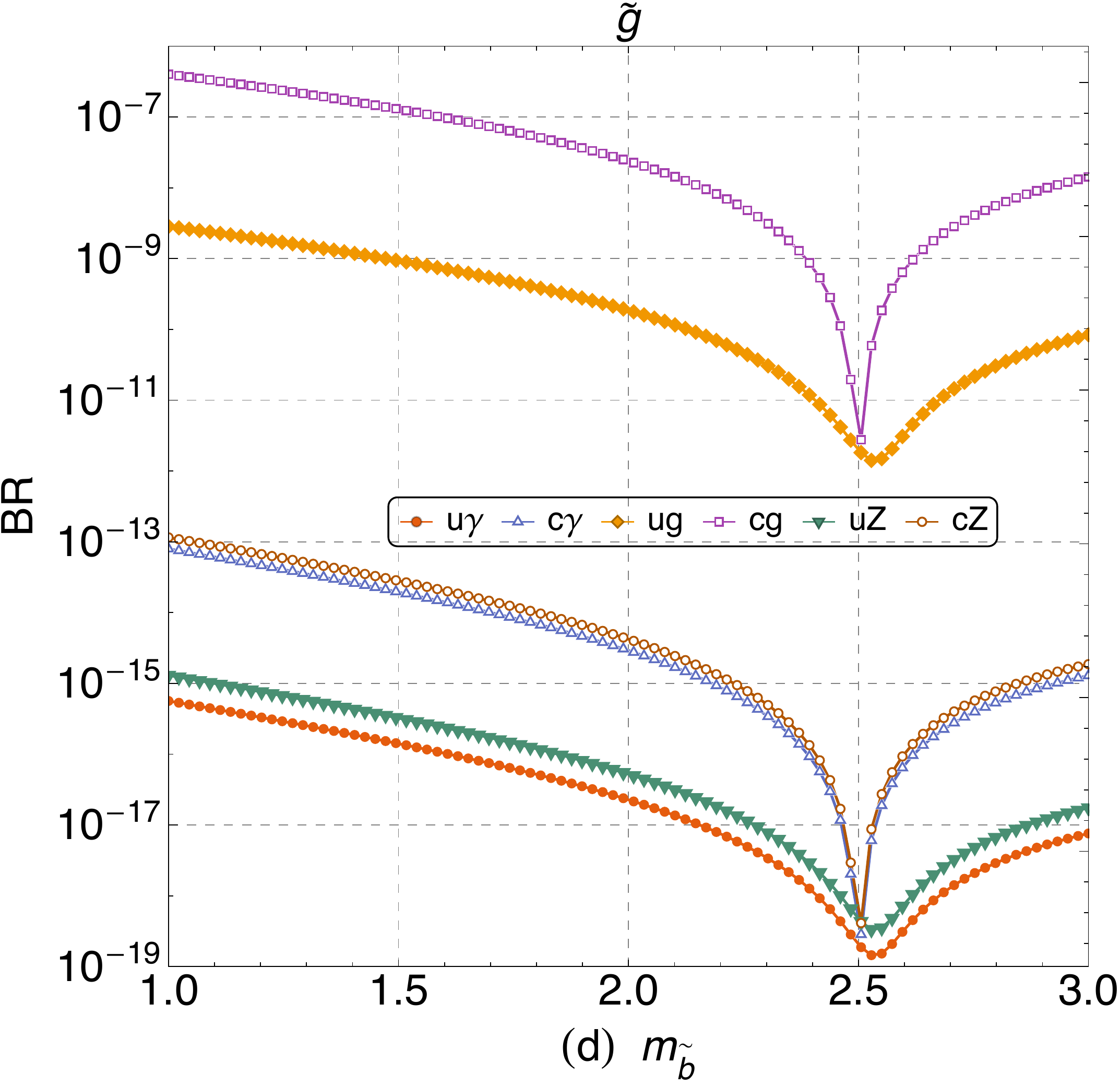}
\caption{The dependence of BR($t\rightarrow qV,qh$) on the third generation squark mass $m_{\tilde{b}}$ in the MRSSM. The values of BR($t\rightarrow qV, qh$) are calculated by including only the specific contributions listed, while setting all other contributions to zero. The squark mass $m_{\tilde{b}}$ is given in TeV.
It is important to note that the contributions from $\chi^{\pm}$ and $\tilde{g}$ to BR($t\rightarrow qh$) are extremely small ($\le 10^{-25}$) and are not shown in the figure.}
\label{figrg1}
\end{figure}
In Fig.\ref{figmb1} (b), we explore another narrow cancellation region around $m_{\tilde{b}}$ = 1 TeV for BR($t\rightarrow qZ$). To explain this phenomenon, we present the predicted BR($t\rightarrow qZ,F^{L/R}_{1Z/2Z}$) values as a function of the third generation squark mass $m_{\tilde{b}}$. In this case, only the contribution from the specific coefficient $F^{L/R}_{1Z/2Z}$ is taken into account. There is a significant decrease at $m_{\tilde{b}}\approx 1.1$ GeV for BR($t\rightarrow qZ, F^{R}_{1Z}$), which explains the leftmost decrease observed in Fig.\ref{figmb1} (a). A more detailed explanation for the dip at $m_{\tilde{b}}\simeq$ 1.1 TeV is done in the discussion of Fig.\ref{figrg1}.

In Fig.\ref{figrg1}, we have independently plotted the contributions to BR($t\rightarrow qV, qh$) from the neutralino $\chi^0$, $\chi$-chargino, $\rho$-chargino, and gluino $\tilde{g}$ as a function of the third generation squark mass $m_{\tilde{b}}$.
For BR($t\rightarrow q\gamma$), we observe that the $\rho$-chargino contribution dominates the predictions, while the contributions from $\chi^0$, $\tilde{g}$, and $\chi$-chargino are less dominant or even negligible.
For BR($t\rightarrow qg$), the leading contribution comes from the $\tilde{g}$, indicating its dominance. The contributions from $\rho$-chargino and $\chi^0$ are less dominant, and the $\chi$-chargino contribution is negligible.
When it comes to BR($t\rightarrow qZ$), both the contributions from $\chi^0$ and $\rho$-chargino are significant and comparable, and they dominate the predictions. The contributions from $\tilde{g}$ and $\chi^{\pm}$ are less dominant or negligible.
The narrow cancellation region observed around $m_{\tilde{b}}$ = 1 TeV for BR($t\rightarrow qZ$) in Fig.\ref{figmb1} and a shift of dip value $m_{\tilde{b}}$ can be explained by the interference between the corrections from the $\rho$-chargino sector and those from the $\chi^0$ sector.
For BR($t\rightarrow qh$), both the contributions from $\chi^0$ and $\rho$-chargino are comparable and dominant in predicting the branching ratio.

Contributions from $\chi^{\pm}$ and $\tilde{g}$ for $t\rightarrow qh$ are very small ($\le 10^{-25}$) and therefore negligible in Fig.\ref{figrg1} (b) and Fig.\ref{figrg1} (d). This is due to the fact that the $\chi^{\pm}$/$\tilde{g}$-quark-squark coupling is either left-handed or right-handed.
To understand this point, let's examine the last diagram in Fig.\ref{Fig1}. Using the $\rho^{\pm}$-mediated diagrams as an example, the factors $F^{L/R}_h$ (e.g. $t\rightarrow uh$) in Eq. (\ref{WCsh}) are proportional to the masses of particles and can be expressed as
\begin{align}
\begin{array}{l}
F^{L}_h\propto C^{u\rho^{\pm}\tilde{d}}_R C^{\ast t\rho^{\pm}\tilde{d}}_R B_0m_{u}m_{\rho^{\pm}}-C^{u\rho^{\pm}\tilde{d}}_L C^{\ast t\rho^{\pm}\tilde{d}}_R B_1m^2_{u}\\[4pt]
\hspace{2.4em}-C^{u\rho^{\pm}\tilde{d}}_R C^{\ast t\rho^{\pm}\tilde{d}}_L B_1m_{u}m_{t}+C^{u\rho^{\pm}\tilde{d}}_L C^{\ast t\rho^{\pm}\tilde{d}}_L B_0 m_{t}m_{\rho^{\pm}},\\[4pt]
F^{R}_h\propto F^{L}_h(L\leftrightarrow R).
\end{array}
\label{FFs}
\end{align}
In the above equations, $C^{u\rho^{\pm}\tilde{d}}_{L/R}$ ($C^{\ast t\rho^{\pm}\tilde{d}}_{L/R}$) represent the left-handed or right-handed couplings for the interaction between $\rho$-chargino and up-type quark/squark. The terms $B_0$ and $B_1$ correspond to the two-point loop integrals. Regarding the $\chi^{\pm}$-mediated diagrams, since the coupling $C^{u\chi^{\pm}\tilde{d}}$ is left-handed, only the term proportional to $B_1m^2_{u}$ for $F^{L}_h$ in Eq. (\ref{FFs}) is nonzero (or $B_1m_{u}m_t$ for $F^{R}_h$). The relationship
\begin{align}
\frac{BR(t\rightarrow qh,\;\chi^{\pm}\;mediated)}{BR(t\rightarrow qh,\;\rho^{\pm}\; mediated)}\propto\frac{m^2_q}{m^2_{\rho^{\pm}}}
\end{align}
can be derived. Since the mass of $\rho$-chargino is at TeV scale, the mass of $u/c$ quark is much smaller than that of $\rho$-chargino. The ratio $\frac{m^2_q}{m^2_{\rho^{\pm}}}$ would be very small. Therefore, the contribution from $\chi$-chargino is negligible when compared to $\rho$-chargino.
A similar comparison can be made between $\chi^{0}$-mediated diagrams and $\tilde{g}$-mediated diagrams.

\begin{figure}[t]
\centering
\includegraphics[width=2.0in]{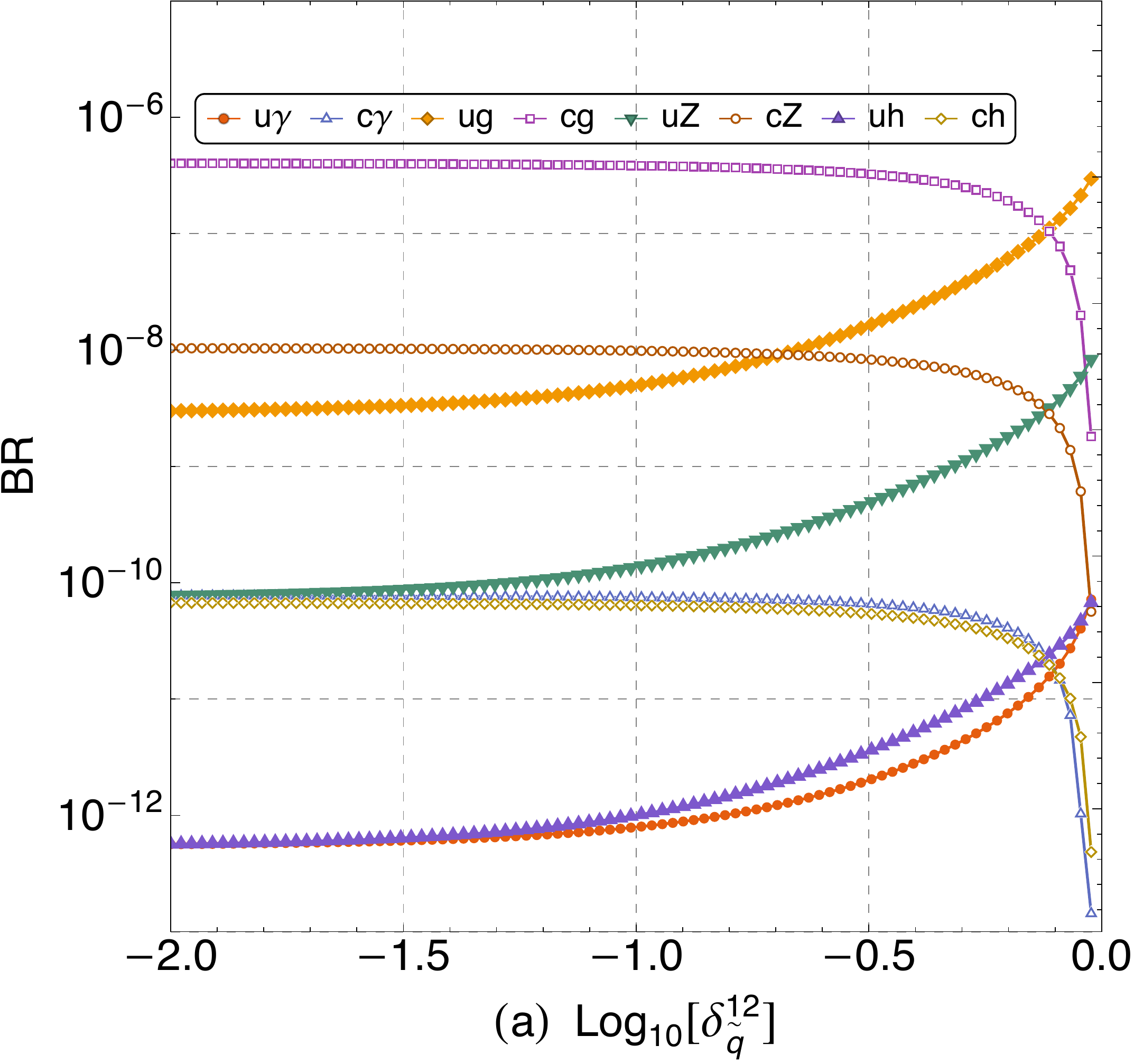}%
\includegraphics[width=2.0in]{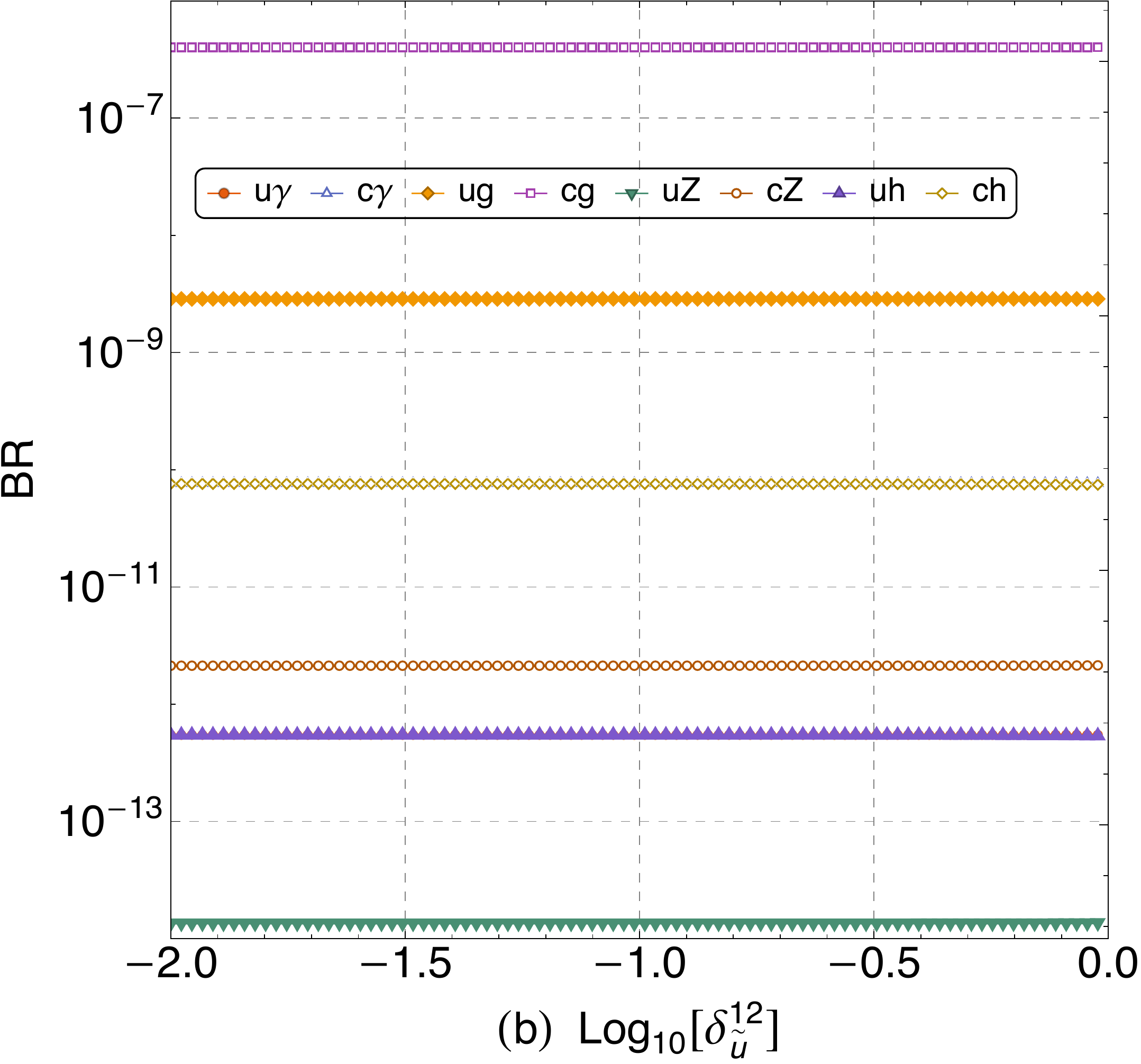}%
\includegraphics[width=2.0in]{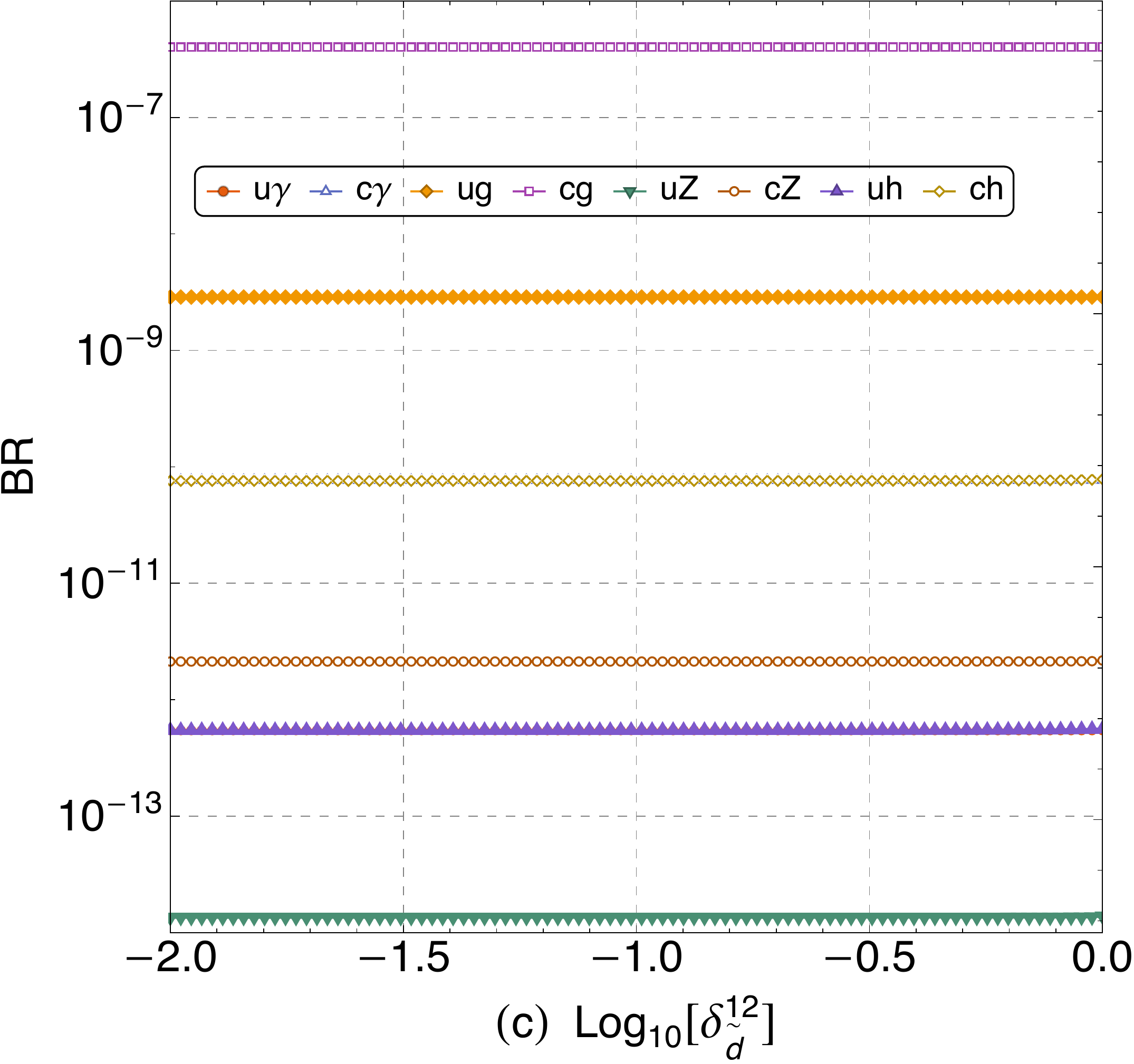}
\includegraphics[width=2.0in]{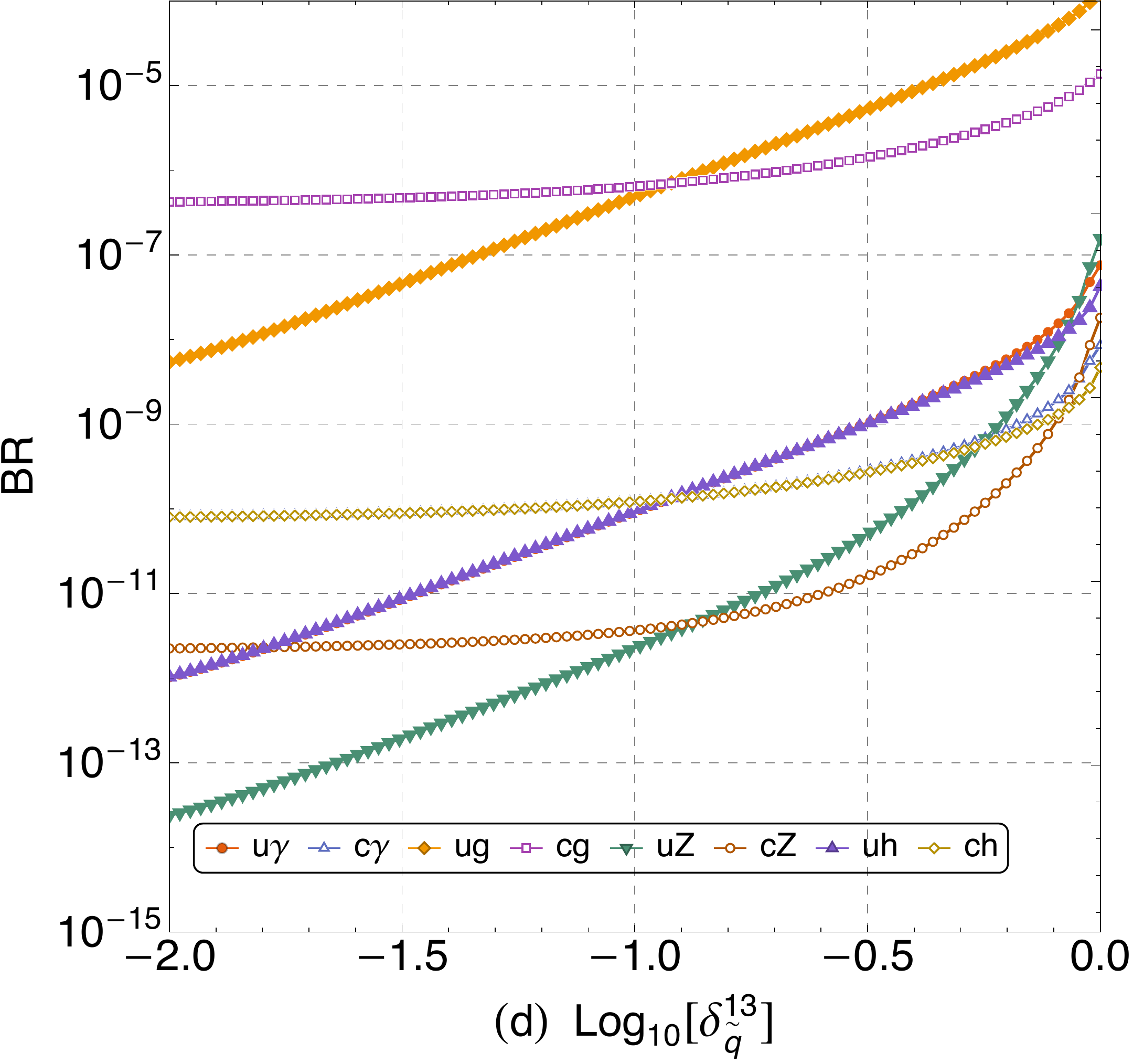}%
\includegraphics[width=2.0in]{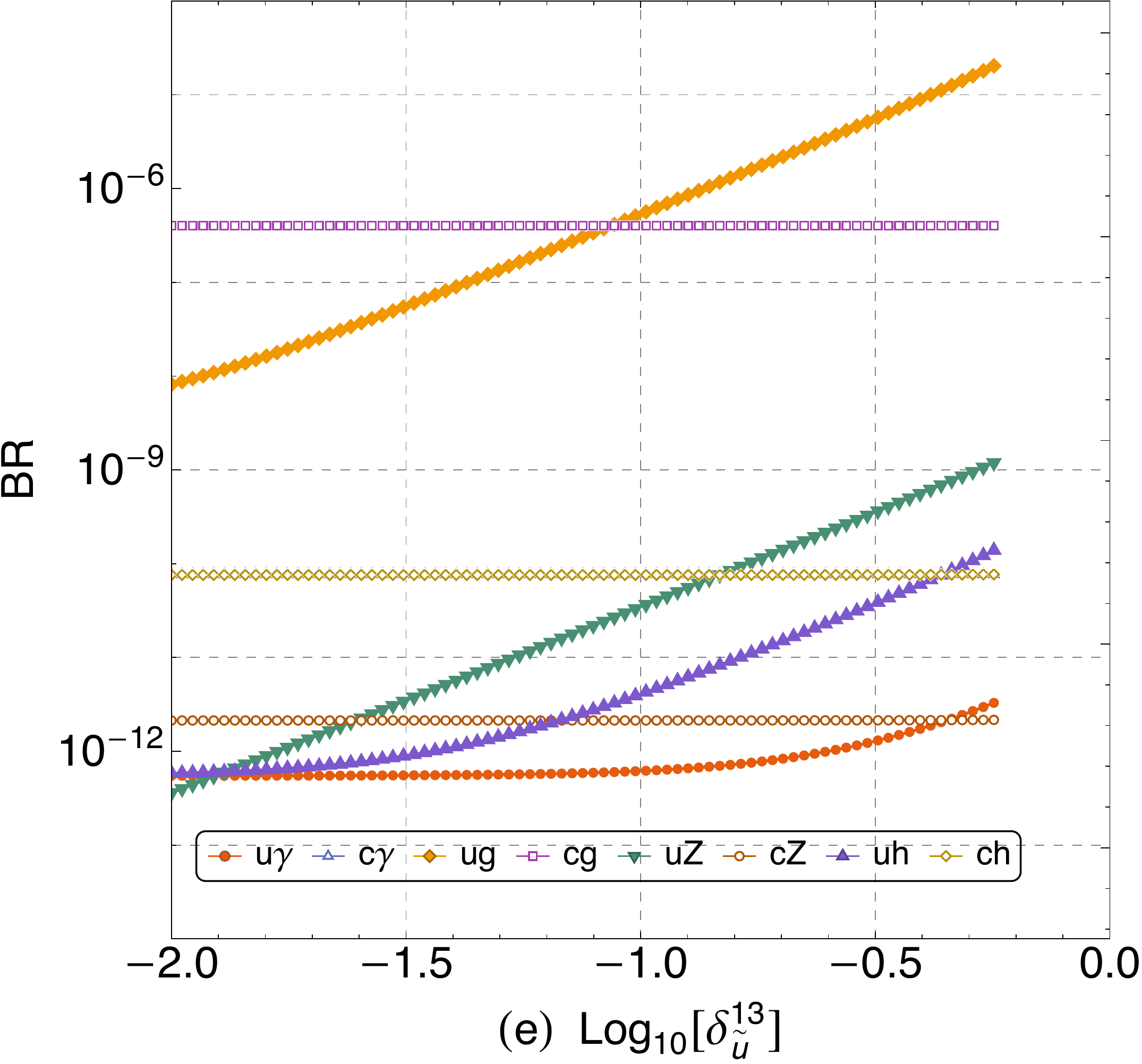}%
\includegraphics[width=2.0in]{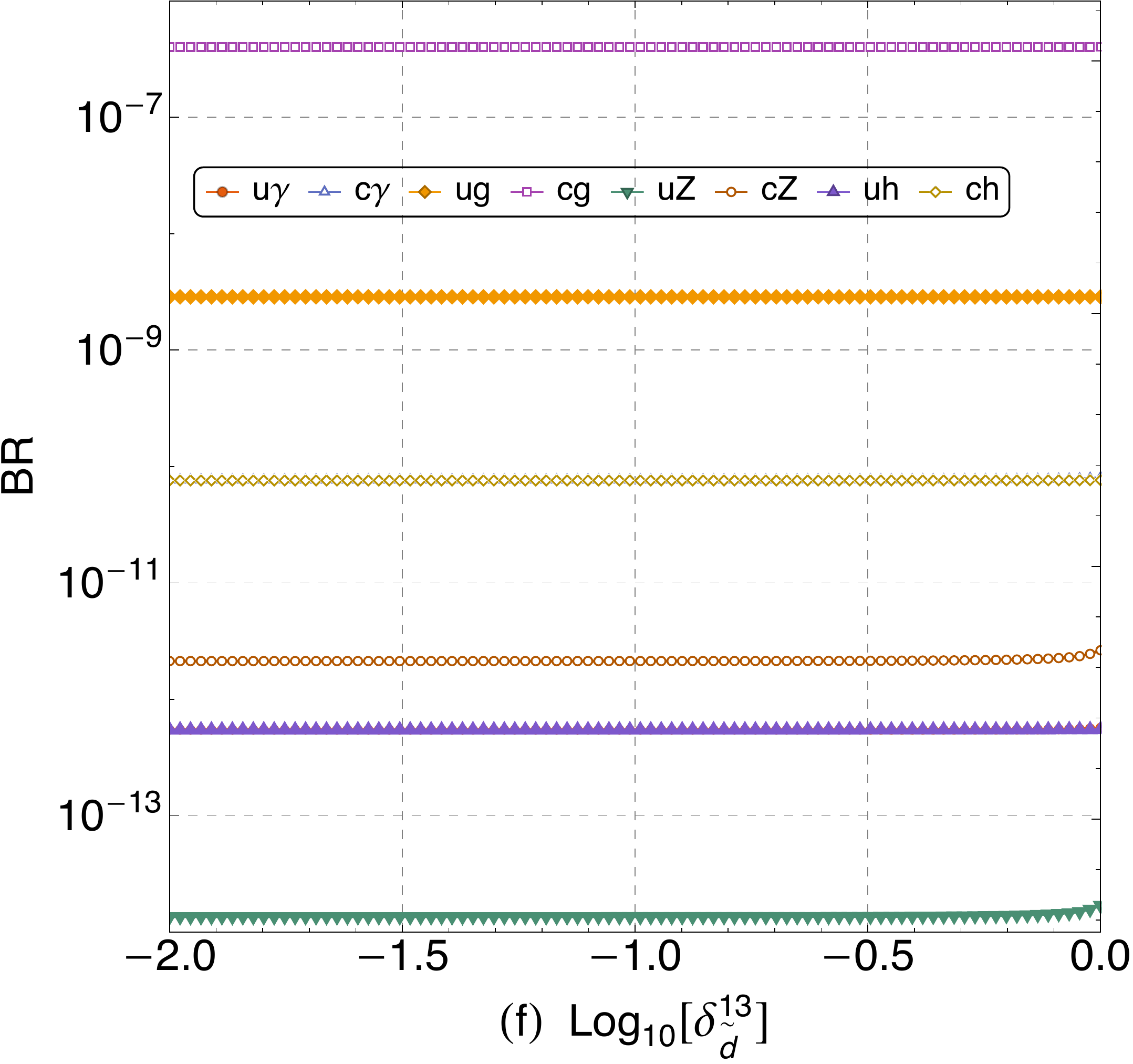}
\includegraphics[width=2.0in]{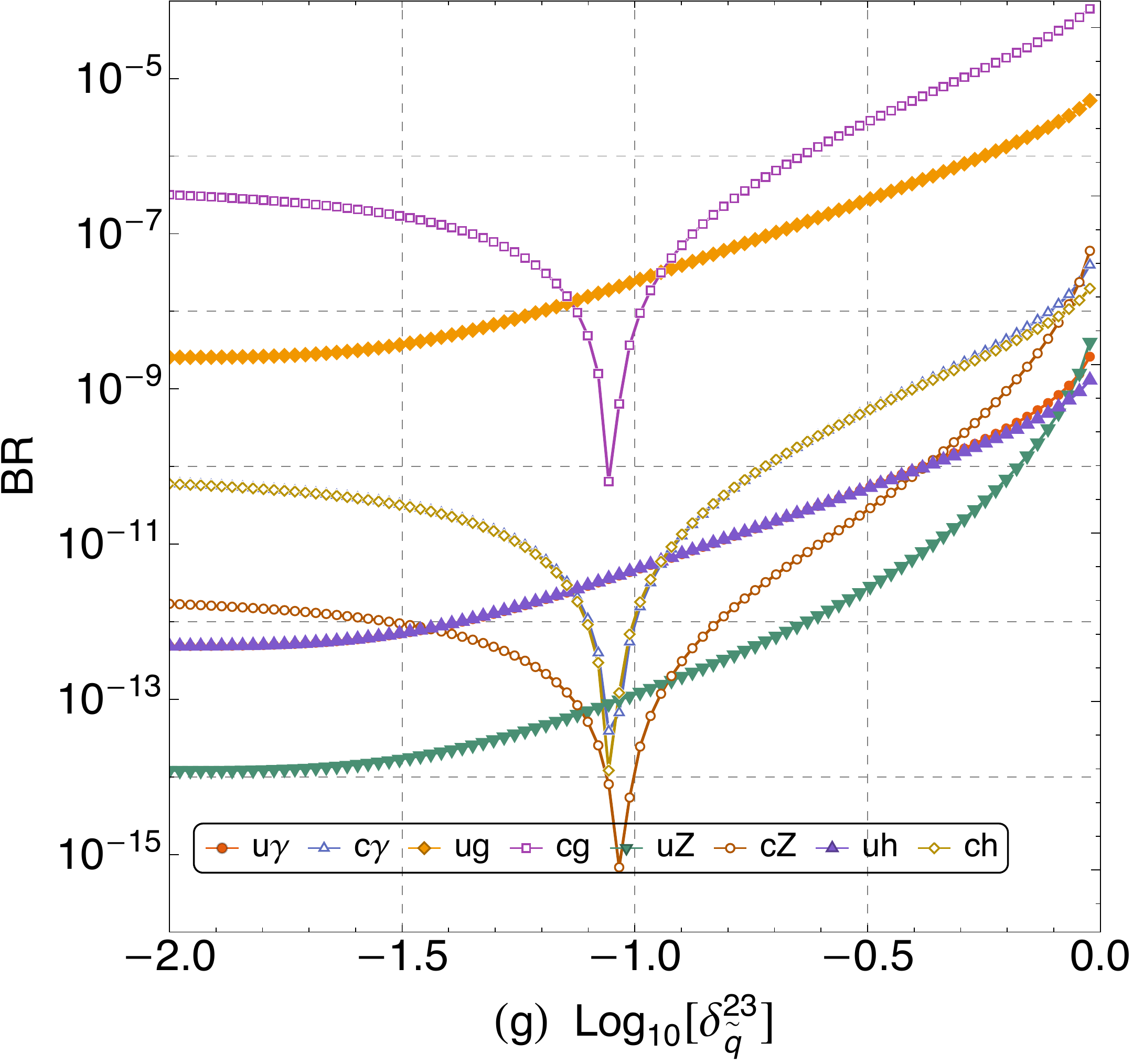}%
\includegraphics[width=2.0in]{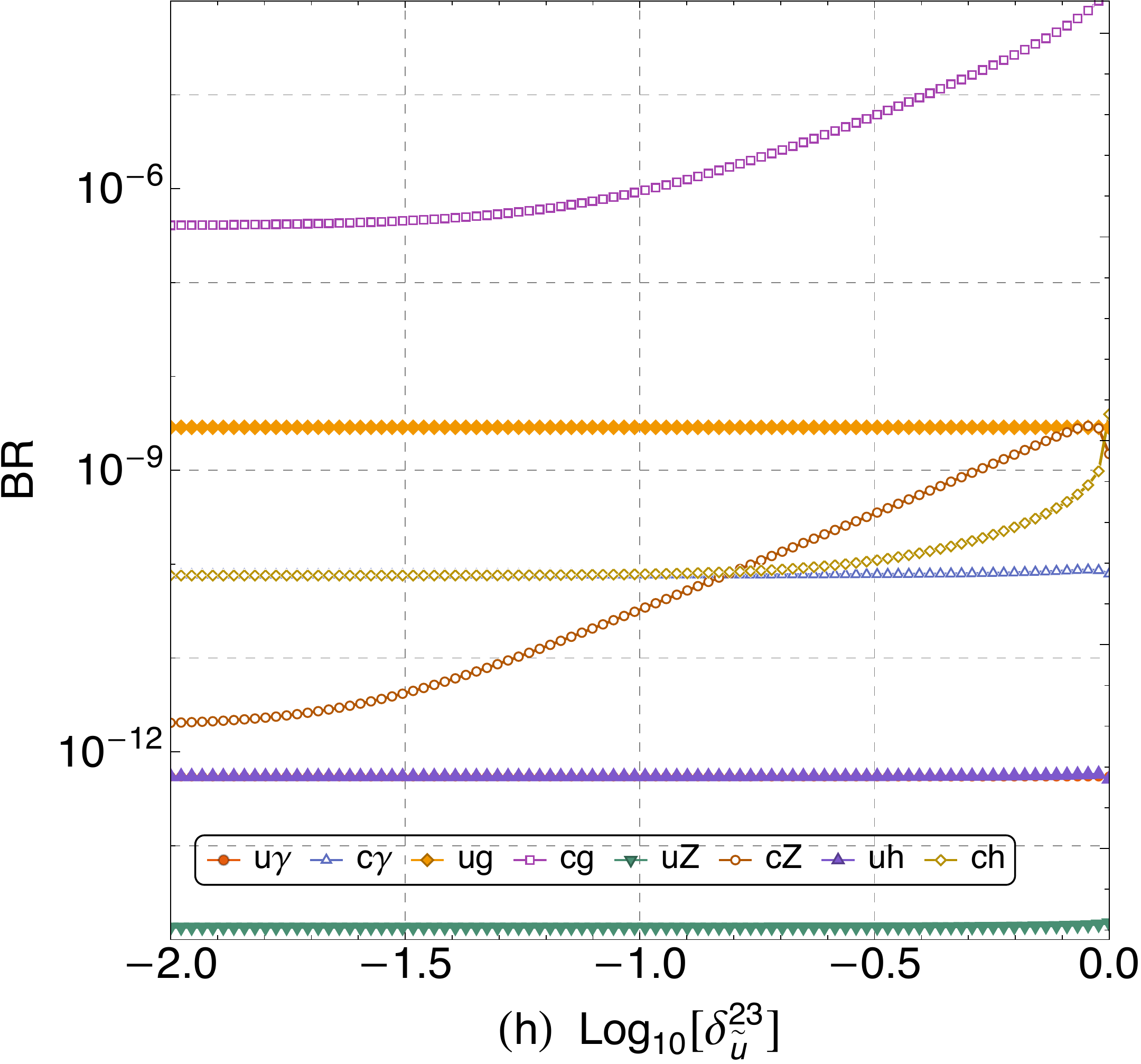}%
\includegraphics[width=2.0in]{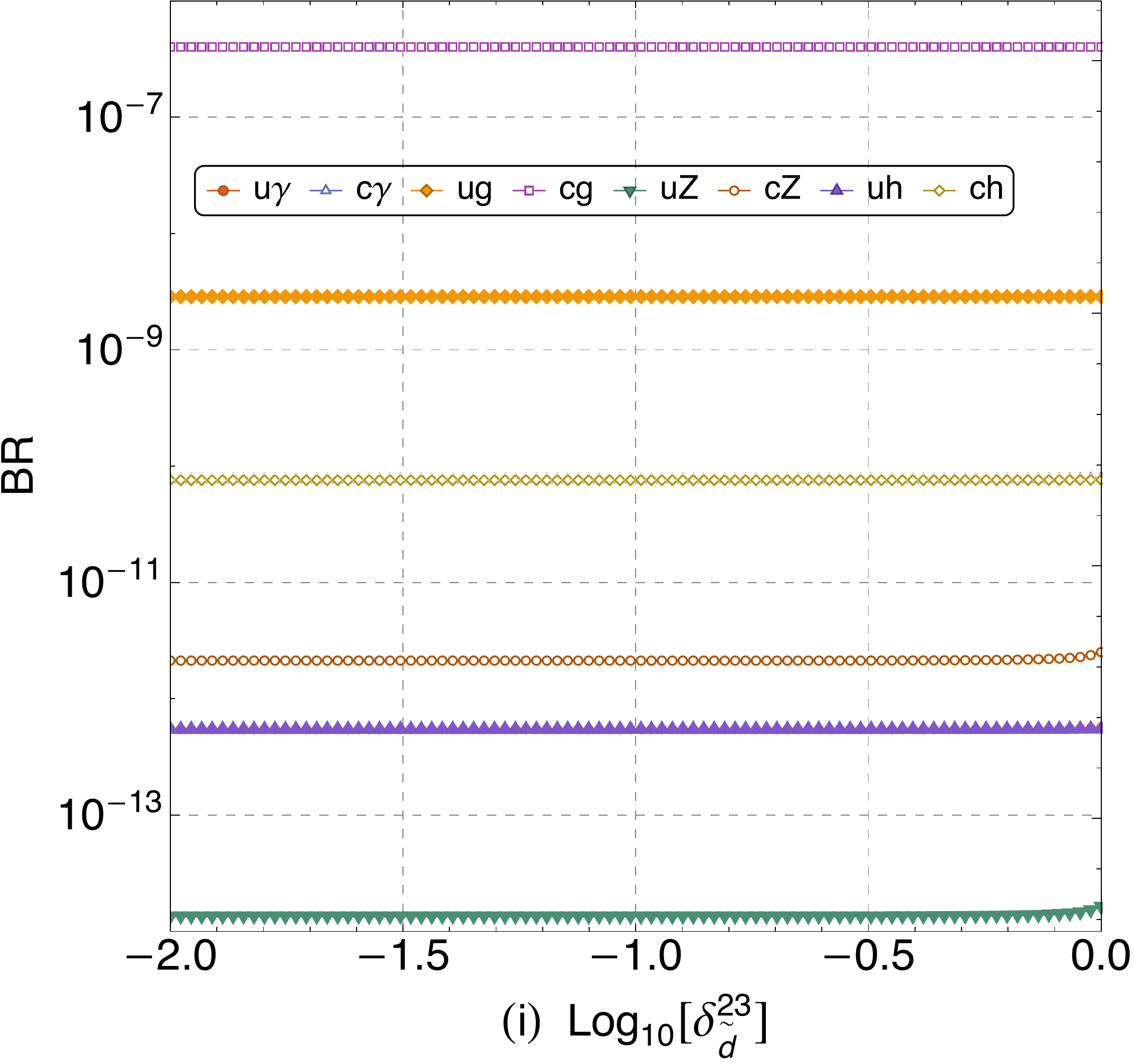}
\caption{The dependence of BR($t\rightarrow qV,qh$) on the base-10 logarithm of the squark flavor mixing parameters $\delta^{IJ}_{\tilde{q}}$,$\delta^{IJ}_{\tilde{u}}$ and $\delta^{IJ}_{\tilde{d}}$.}
\label{figB1d123}
\end{figure}
In several articles \cite{Bejar2001, Guasch1999, Cao2007, Liu2004, Delepine2004, Dedes2014, Frank2005}, it has been discussed that the off-diagonal entries of the soft-breaking terms $m_{\tilde{q}}^{2}$, $m_{\tilde{u}}^{2}$, and $m_{\tilde{d}}^{2}$ can greatly influence the predicted BR($t\rightarrow qV, qh$) in various supersymmetric standard models. In the following analysis, we will investigate the impact of these squark flavor mixing parameters on BR($t\rightarrow qV, qh$) in the MRSSM.
The off-diagonal elements of the squark mass matrices $m_{\tilde{q}}^{2}$, $m_{\tilde{u}}^{2}$, and $m_{\tilde{d}}^{2}$ can be parameterized using the mass insertion method, where $(m^{2}_{X})_{IJ}=\delta ^{IJ}_{X}\sqrt{(m^{2}_{X})_{II}(m^{2}_{X})_{JJ}}$ for $X=\tilde{q},\tilde{u},\tilde{d}$, and $I, J=1,2,3$. It should be noted that the default values for the off-diagonal entries of the squark mass matrices $m^2_{\tilde{q}}$, $m^2_{\tilde{u}}$, and $m^2_{\tilde{d}}$ in Eq. (\ref{N1}) are set to zero.

In Fig.\ref{figB1d123}, the predictions for BR($t\rightarrow qV, qh$) are presented as a function of the mass insertion parameters $\delta^{IJ}_{\tilde{q}(\tilde{u},\tilde{d})}$. In each plot, only the indicated $\delta^{IJ}_{\tilde{q}(\tilde{u},\tilde{d})}$ is varied, while all other mass insertions are set to zero. The predictions are constrained by experimental data for BR($\bar{B}\rightarrow X_s\gamma$), BR($B^0_s\rightarrow \mu^+\mu^-$), and BR($B^0_d\rightarrow \mu^+\mu^-$).
From the results, it can be observed that varying the parameters $\delta^{12}_{\tilde{u}}$, $\delta^{23}_{\tilde{u}}$, $\delta^{12}_{\tilde{d}}$, $\delta^{13}_{\tilde{d}}$, and $\delta^{23}_{\tilde{d}}$ has a minimal effect on the predictions of BR($t\rightarrow uV, uh$). Similarly, varying the parameters $\delta^{12}_{\tilde{u}}$, $\delta^{13}_{\tilde{u}}$, $\delta^{12}_{\tilde{d}}$, $\delta^{13}_{\tilde{d}}$, and $\delta^{23}_{\tilde{d}}$ has a small effect on the predictions of BR($t\rightarrow cV, ch$). In both cases, the predictions fall within a narrow band.
The predictions of BR($t\rightarrow uV, uh$) increase with an increase in the values of $\delta^{12}_{\tilde{q}}$, $\delta^{13}_{\tilde{q}}$, $\delta^{23}_{\tilde{q}}$, and $\delta^{13}_{\tilde{u}}$. Moreover, the predictions of BR($t\rightarrow ug$) can approach the current experimental bound, while the predictions for other processes remain several orders of magnitude below the experimental bounds.
For the predictions of BR($t\rightarrow cV, ch$), they increase with an increase in the values of $\delta^{13}_{\tilde{q}}$ and $\delta^{23}_{\tilde{u}}$, but decrease with an increase in $\delta^{12}_{\tilde{q}}$. Notably, a significant decrease can be observed at $Log_{10}[\delta^{23}_{\tilde{q}}]\sim-1.0$ for BR($t\rightarrow cV, ch$), which arises from destructive interference. This behavior in the MRSSM is similar to that observed in the MSSM \cite{Cao2007} and the left-right supersymmetric model \cite{Frank2005}.

\begin{figure}[h]
\centering
\includegraphics[width=0.5\columnwidth]{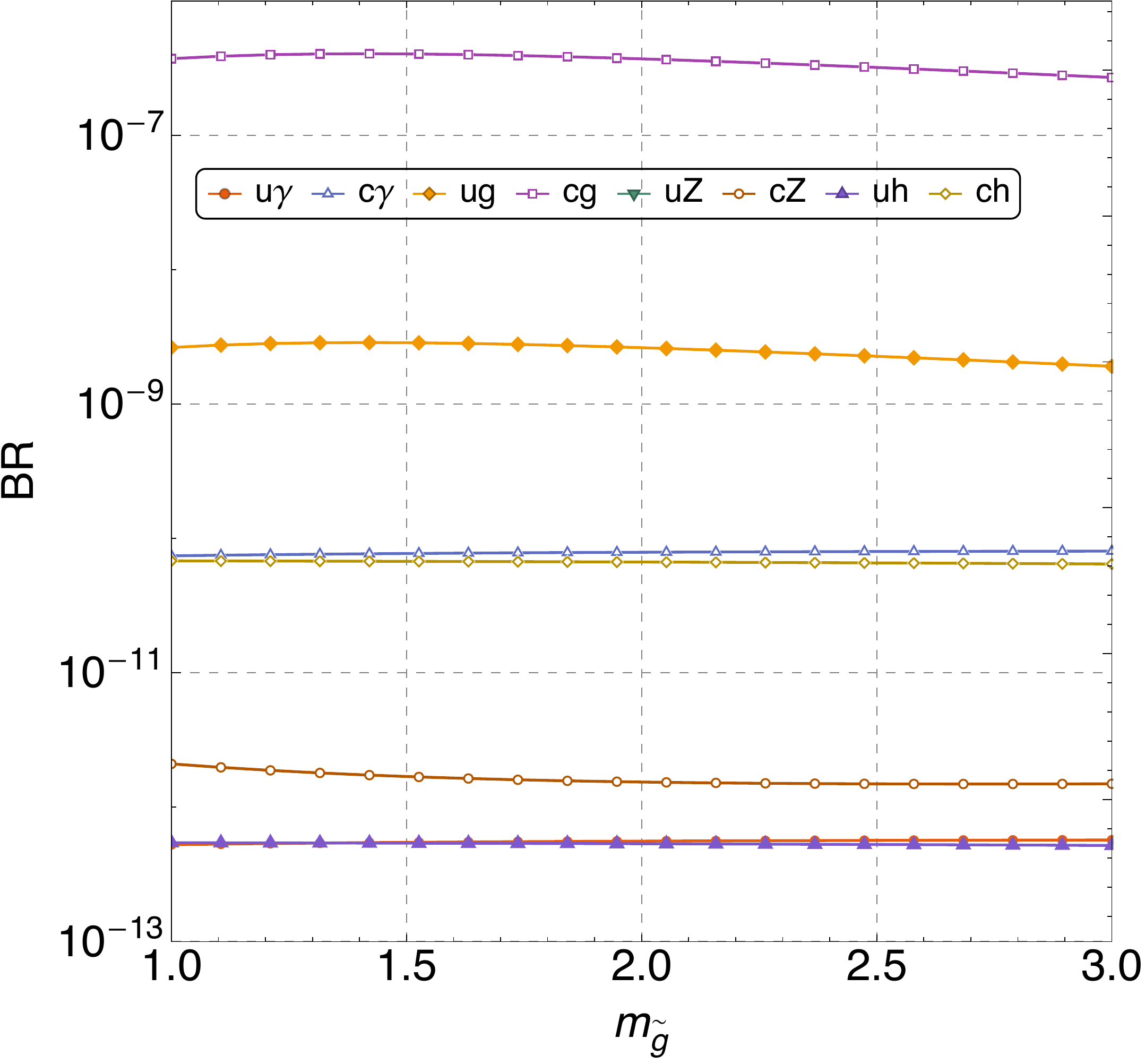}
\caption{The dependence of BR($t\rightarrow qV,qh$) on the gluino mass $m_{\tilde{g}}$. $m_{\tilde{g}}$ is given in TeV.}
\label{figmg}
\end{figure}

In Fig.\ref{figmg}, the predictions of BR($t\rightarrow qV, qh$) are plotted as a function of the gluino mass $m_{\tilde{g}}$, using the values given in Eq. (\ref{N1}). The results indicate that varying the gluino mass $m_{\tilde{g}}$ has a minimal effect on the predictions of BR($t\rightarrow qV, qh$), as they fall within a narrow band.
This behavior is distinct from what is typically observed in the MSSM, where the predictions of BR($t\rightarrow qV, qh$) tend to decrease rapidly as the gluino mass increases \cite{Li1994, Guasch1999, Cao2007}. Therefore, the MRSSM exhibits a different behavior in terms of the influence of the gluino mass on the branching ratios.

\section{Conclusions\label{sec4}}

In this work, we analyze the FCNC processes $t\rightarrow qV, qh$ within the framework of the minimal R-symmetric supersymmetric standard model, while considering the constraints imposed by experimental data on the parameter space.
The MRSSM is characterized by the absence of Majorana gaugino masses, $\mu$ term, $A$ terms, and left-right squark and slepton mass mixings due to the presence of R-symmetry. This absence of certain terms sets the MRSSM apart from the MSSM and leads to different predictions for BR($t\rightarrow qV, qh$). For example, the trilinear $A^{23}_u$-terms are important for $t\rightarrow ch$ in the MSSM.

In our analysis, we investigate the dependence of BR($t\rightarrow qV, qh$) on several parameters, including the ratio tan$\beta$, the third generation squark mass $m_{\tilde b}$, and the squark flavor mixing parameters $\delta^{IJ}_{\tilde{q},\tilde{u},\tilde{d}}$.
Additionally, we consider the influence of the gluino mass $m_{\tilde{g}}$ on the branching ratios.
Furthermore, we independently study the contributions to the BR($t\rightarrow qV, qh$) from various supersymmetric particles, allowing us to identify the individual effects of these particles on the process. It is explained that the dips observed in Fig.\ref{figtanb1} and Fig.\ref{figmb1} arise from the contributions of different supersymmetric particles. Isolating the leading contributions analytically can be highly beneficial when dealing with so complicated models. One, for example, may use the Flavour Expansion Theorem \cite{Dedes2015} to accomplish this goal.

At the benchmark points, the predicted BR($t\rightarrow q\gamma,qZ,qh$) are in the range of $\mathcal O$$(10^{-11}$$\sim$$10^{-14})$, which are only a few orders of magnitude above the SM prediction. By considering the effects of parameters such as $\delta^{13}_{\tilde{q}}$ or $\delta^{23}_{\tilde{u}}$, the predicted BR($t\rightarrow q\gamma, qZ, qh$) can be enhanced to $\mathcal O(10^{-9}\sim10^{-11})$, which are four or five orders of magnitude below the estimated branching ratios at future high-luminosity colliders like HL-LHC, HE-LHC and FCC-hh \cite{Aguilar2017, YJZhang}.

The decay final states $t \rightarrow ug, cg$ are challenging to detect directly in hadron colliders due to the large backgrounds and the difficulty in identifying the final state particles. Instead, the inverse processes, $ug \rightarrow t$ and $cg \rightarrow t$, have been utilized by experiments such as CMS \cite{Khachatryan2017} and ATLAS \cite{ATLAS2016} to place limits on the corresponding branching ratios. In future projects, such as HL-LHC \cite{CMSqg} and FCC-hh \cite{Oyulmaz}, a sensitivity of the order of $10^{-6}\sim 10^{-7}$ for BR($t\rightarrow ug, cg$) would be achievable.
At the benchmark points, the predicted BR($t\rightarrow ug, cg$) in the MRSSM are at the level of $\mathcal O(10^{-7})$ and $\mathcal O(10^{-9})$. By considering the effects of parameters such as $\delta^{13}_{\tilde{q}}$, $\delta^{23}_{\tilde{q}}$, $\delta^{13}_{\tilde{u}}$, or $\delta^{23}_{\tilde{u}}$, the predicted BR($t\rightarrow ug, cg$) can be enhanced to the level of $\mathcal O(10^{-5}\sim10^{-6})$. These enhanced values can potentially be tested at the HL-LHC and FCC-hh.

\begin{acknowledgments}
\indent\indent
This work has been supported partly by the National Natural Science Foundation of China (NNSFC) under Grant No.11905002, the Natural Science Foundation of Hebei Province under Grants No.A2022104001 and No.A2022201017, the Natural Science Basic Research Program of Shaanxi (Program No. 2023-JC-YB-072), the youth top-notch talent support program of the Hebei Province and the Foundation of Baoding University under Grant No. 2019Z01.
\end{acknowledgments}


\end{document}